\documentclass[12pt, draftclsnofoot, onecolumn]{IEEEtran}
\usepackage{amssymb}
\usepackage{amsmath}
\usepackage{cite}
\usepackage{url}
\usepackage{xcolor}
\usepackage{cite,graphicx,amsmath,amssymb}
\usepackage{subfigure}
\usepackage{citesort}
\usepackage{fancyhdr}
\usepackage{mdwmath}
\usepackage{mdwtab}
\usepackage{caption}
\usepackage{amsthm}
\usepackage{setspace}
\usepackage{algorithm}
\usepackage{algorithmic}
\usepackage{makecell}
\usepackage{diagbox}

\newtheorem{theorem}{Theorem}

\newtheorem{lemma}{Lemma}

\newtheorem{corollary}{Corollary}

\allowdisplaybreaks
\makeatletter
\def\ScaleIfNeeded{%
\ifdim\Gin@nat@width>\linewidth \linewidth \else \Gin@nat@width
\fi } \makeatother
\begin{document}
\title{Exploiting Semantic Communication for Non-Orthogonal Multiple Access}
\author{

Xidong~Mu,~\IEEEmembership{Member,~IEEE,} and
       Yuanwei~Liu,~\IEEEmembership{Senior Member,~IEEE,}

\thanks{Xidong Mu and Yuanwei Liu are with the School of Electronic Engineering and Computer Science, Queen Mary University of London, London E1 4NS, U.K. (email: xidong.mu@qmul.ac.uk; yuanwei.liu@qmul.ac.uk).}

}

\maketitle
\vspace{-1.8cm}
\begin{abstract}
A novel semantics-empowered two-user uplink non-orthogonal multiple access (NOMA) framework is proposed for resource efficiency enhancement. More particularly, a secondary far user (F-user) employs the semantic communication (SemCom) while a primary near user (N-user) employs the conventional bit-based communication (BitCom). The fundamental performance limit, namely \emph{semantic-versus-bit (SvB) rate region}, of the proposed semantics-empowered NOMA framework is characterized. The \emph{equivalent} SvB rate region achieved by the conventional BitCom-based NOMA is provided as the baseline scheme. It unveils that, compared to BitCom, SemCom can significantly improve the F-user's performance when its permitted transmit power is strictly capped, but may perform worse when its permitted transmit power is high. Guided by this result, the proposed semantics-empowered NOMA framework is investigated over fading channels. An \emph{opportunistic} SemCom and BitCom scheme is proposed, which enables the secondary F-user to participate in NOMA via the most suitable communication method at each fading state, thus striking a good tradeoff between its own achieved performance and the interference imposed to the primary N-user. Two scenarios are considered for employing the opportunistic scheme, namely on-off resource management and continuous resource management. For each scenario, the optimal communication policy over fading channels is derived for maximizing the ergodic semantic rate achieved at the secondary F-user, subject to the minimum ergodic bit rate constraint of the primary N-user. Numerical results show that: 1) The proposed opportunistic scheme in both scenarios can achieve higher communication performance for NOMA than the baseline schemes merely employing SemCom or BitCom; 2) SemCom can better guarantee the performance of the F-user admitted in NOMA than BitCom when the communication requirement of the primary N-user is high; and 3) The continuous power control at the F-user is necessary for ensuring high performance over fading channels, while the on-off time scheduling is sufficient.
\end{abstract}

\section{Introduction}
Thanks to the rapid development of wireless communications, people's life and work style have been significantly improved. Compared to the earliest wireless communication technology in the 1920s, the current 5G wireless communication system enables people to communicate with each other throughout speech, text, and video in a ubiquitous manner. Reviewing the historical development of wireless communications, the Shannon classical information theory~\cite{Shannon2} provides the fundamental rate limit for realizing reliable communications. In the Shannon paradigm, the design metric is to ensure the bit sequences converted from the source can be accurately transmitted to the destination, where the intrinsic meaning of the source is put aside during the transmission. Guided by the Shannon paradigm, ``engineering'' problems can be formulated considering generalized performance metrics, such as symbol-error rate (SER) and bit-error rate (BER), which thus receive growing research efforts in the past a few decades. Despite a series of sophisticated communication technologies~\cite{6736761,6852102,6736746} have been proposed for approaching or achieving the Shannon rate limit, the communication requirements have not been completely fulfilled, which calls for the development of new communication technologies.

Semantic communication (SemCom) has received significant attention in recent years and is regarded as a promising technology in future wireless networks~\cite{Gunduz,Ping,Qin}. The main idea of SemCom is to only transmit the key information, that are able to deliver the desired meaning/actions/goals to the destination, and safely remove the information irrelevant to the specific task without causing any performance degradation~\cite{Qin}. Compared to the Shannon paradigm, where the bit sequence representing the entire source has to be always transmitted, the size of source to be transmitted in SemCom can be greatly compressed in terms of the specific tasks, thus reducing the required power and spectrum resources. It is worth mentioning that SemCom matches well with the developing trend of wireless communications. This is because, on the one hand, the scale of wireless networks is still explosively growing without limitations. For example, it is predicted that the global amount of data generated by network nodes will increase to 175 zetta-bytes in 2025~\cite{ICD}. For satisfying such stringent data rate and connectivity requirements, the relentless increase of the transceivers' size and operating frequencies in the Shannon paradigm would result in extremely high energy consumption and hardware cost, especially in the current Post-Moore law era. In this case, SemCom comes to the rescue given its sustainable feature. On the other hand, the revolutionary killer applications (e.g., Telemedicine, Smart Cities, Metaverse, etc.) in future wireless networks have to support ``human-to-machine (H2M)'' and ``machine-to-machine (M2M)'' communications~\cite{H2M,M2M}, where the main purpose is to make the receiver understand the intrinsic meaning of the source and take the right actions accordingly, i.e., following the salient principle of SemCom. As a result, considering the need of an intelligent and sustainable communication paradigm, the development of SemCom is no longer an option but a necessity.

\subsection{Prior Works}
The concept of SemCom was first introduced by Weaver and Shannon in 1949~\cite{Shannon2}, where communication problems were classified into three levels, a technical level, a semantic level, and an effectiveness level. SemCom aims to address the later two levels of communication problems. Since then, growing research efforts have been devoted into establishing theoretical frameworks for SemCom. As a first step, the authors of \cite{Carnap} presented a definition of semantic information, where the semantic entropy of a sentence under a given language system is measured by employing the logical probabilities of the sentences. The author of \cite{Floridi} proposed a quantitative theory of strongly semantic information, where the relative information of the considered sentence with respect to the given reference sentence is measured by the truth-value instead of its own probability distributions. Based on the results of \cite{Carnap}, the authors of \cite{6004632} proposed a general framework for measuring the semantic information in communications, where the semantic noise and semantic channel are defined as well as the semantic channel capacity is obtained by extending the Shannon's channel coding theorem. In recent years, the authors of \cite{9518240} conceived a general rate-distortion framework for characterizing the semantic information, where the target information source is modelled by an intrinsic state and an extrinsic observation for the receiver to infer. The authors of \cite{9814642} proposed a novel SemCom framework, where a semantic communication (SC) layer is added on the top of the current technical communication (TC) layer. Throughout the beneficial interaction between the SC and TC layers, the ultimate communication efficiency can be improved. 

Besides the theoretical studies of SemCom, some researchers began to investigate the implementation of SemCom with the aid of machine learning tools. Current research contributions can be broadly classified into two categories. In the first category, SemCom is designed to improve the transmission performance of different types of data, such as text, speech, and image. In particular, the authors of \cite{8461983} proposed a joint source and channel coding (JSCC) text transmission approach, where the deep learning (DL) based encoder and decoder were developed for processing semantic information contained in the sentence. The authors of \cite{9398576} further proposed an advanced DL based SemCom tool (termed as DeepSC) for text transmission, which outperforms conventional transmission methods in low and moderate signal-to-noise ratio (SNR) regime. Based on~\cite{9398576}, the authors of \cite{9763856} designed a performance metric, namely semantic rate, for characterizing the performance of SemCom. The authors of \cite{9791409} further improved the performance of semantic text transmission by integrating the semantic coding with conventional Reed-Solomon channel coding and hybrid automatic repeat request. For image transmission, the authors of \cite{8723589} developed a DL-based JSCC approach, where DL is adopted to encode the image pixel values into the transmitted symbols. It shows that the proposed approach outperforms the conventional image compression and encoding methods in the low SNR regime. In the other category, SemCom is employed for realizing the specific task execution, such as image classification, machine translation, and visual question answering. For example, the authors of \cite{9837474} developed task-oriented SemCom in multi-device cooperative edge inference systems for carrying out multi-view image classification tasks and multi-view object recognition tasks. The authors of \cite{9796572} investigated the task-oriented image transmission for achieving image classification at unmanned aerial vehicles. The authors of \cite{9830752} studied the task-oriented multi-user SemCom in the single-modal case for image retrieval and machine translation as well as the multi-modal case for visual question answering. 
\subsection{Motivations and Contributions}
Compared to conventional bit-level communications under the Shannon paradigm (termed as ``BitCom'' in this paper), SemCom provides a new communication paradigm for facilitating intelligent information exchange. Existing research contributions reveals that SemCom is promising to be used when the SNR is low and/or the available wireless resource is limited~\cite{Qin,9398576}. In other words, for achieving the same communication performance, SemCom generally requires less power/bandwidth than BitCom. Sparked by the above salient advantages of SemCom, it is natural to investigate the application of SemCom in other promising communication technologies for further performance enhancement. Among them,
non-orthogonal multiple access (NOMA) is an important technology for supporting massive connectivity and improving spectrum efficiency in future wireless networks. The key idea of NOMA is to accommodate multiple users over the same resource block, where successive interference cancelation (SIC) is employed to deal with the resulting interference among users~\cite{Liu2017}. However, given the sequential nature of SIC, the user whose signal is decoded at an earlier stage always suffer from stronger interference than the user whose signal is decoded at the later stage~\cite{Ding}. This incurs the so-called ``early-late'' rate disparity issue in NOMA, where the achieved communication rate of NOMA users varies significantly with respect to their assigned SIC decoding order. Bearing this issue in mind, previous works~\cite{Ding,SWIPT,9679390} proposed a series of efficient strategies for user paring and SIC decoding order design, such that the users in NOMA can be served in their own right position. However, the ``early-late'' rate disparity issue remains unsolved.

This paper revisits the ``early-late'' rate disparity issue in NOMA by exploiting the new emerging technology of SemCom. In particular, a fundamental two-user uplink NOMA communication system is considered, which consists of a primary near user (N-user), a secondary far user (F-user), and an access point (AP). The F-user tries to reuse the resource block occupied by the N-user for uploading information to the AP. In terms of the channel condition difference of the two users, the AP will first decode the primary N-user's signal at the interference caused by the admitted secondary F-user's signal, and then decodes the F-user's signal after SIC. Conventionally, the secondary F-user can only be admitted to the channel when at least one user has a small communication requirement, i.e., the ``early-late'' rate disparity issue. To demonstrate it, on the one hand, when the F-user has a small communication requirement, it will not impose a high interference level to the N-user at the first stage of SIC. On the other hand, when the N-user has a small communication requirement, it can tolerate a high interference level such that the F-user can be admitted. It can be observed that it is generally impossible to make a high communication requirement satisfied at both users in conventional NOMA. To address this issue, this paper aims to investigate how to exploit SemCom to improve the performance for NOMA. The main motivation is that SemCom enables users to achieve a comparable performance as BitCom but only consuming a sufficiently low transmit power, i.e., causing less interference to other users. To the best of the authors' knowledge, this is the first work to investigate the potentials of semantics-empowered NOMA transmission.

The main contributions of this paper can be summarized as follows:
\begin{itemize}
  \item We propose a novel semantics-empowered two-user uplink NOMA framework, where one primary N-user communicates with the AP via BitCom and one secondary F-user aims to reuse the N-user's resource block to upload information with the aid of SemCom. To further control the interference caused to the N-user by admitting the F-user, we assume that the total transmission time is partially allocated to the F-user. Based on the proposed framework, we characterize the achieved semantic-versus-bit (SvB) rate region, which is compared with the equivalent one achieved by conventional BitCom-based NOMA. It shows that SemCom can significantly improve the F-user's performance without degrading the N-user's performance, thus alleviating the ``early-late'' rate disparity issue. However, it also reveals that SemCom would perform worse than BitCom when the F-user is permitted to use a high transmit power.
  \item We further investigate the proposed semantics-empowered NOMA framework over fading channels. Inspired by the obtained performance comparison, we propose an opportunistic SemCom and BitCom scheme for enabling the F-user to choose the efficient communication strategy when being admitted into the N-user's channel at each fading state. In terms of the adjustment feature of power control and time scheduling, two scenarios are considered, namely 1) on-off resource management scenario with a constant transmit power; and 2) continuous resource management scenario with both the instantaneous peak and long-term average transmit power constraints. 
  \item We formulate the optimization problem in each scenario for maximizing the ergodic semantic rate of the F-user, subject to the minimum ergodic bit rate constraint of the N-user. As the formulated problem satisfies the time-sharing condition, the optimal opportunistic communication policy at the F-user in each scenario is obtained by employing the Lagrangian dual method.
  \item Our numerical results show that the proposed opportunistic scheme achieves the best performance compared to the baseline schemes relying only on BitCom or SemCom. For NOMA, the employment of SemCom at the secondary F-user can achieve significantly higher performance than BitCom when the primary N-user has a high communication requirement. It also reveals that the continuous power control at the F-user is more important than the continuous time scheduling for improving the communication performance over fading channels.   
\end{itemize}
\subsection{Organization}
The rest of this paper is organized as follows: Section II proposes the semantics-empowered two-user uplink NOMA framework and characterizes the corresponding SvB rate region, which is compared with the conventional BitCom-based NOMA. Section III investigates the proposed framework over fading channels, where an opportunistic SemCom and BitCom scheme is proposed under two scenarios and the corresponding ergodic semantic rate maximization problem is formulated. The optimal communication policy in each scenario for using the opportunistic scheme is derived by invoking the Lagrangian dual method in Section IV. Section V provides numerical results to verify the effectiveness of the proposed opportunistic scheme. Finally, Section VI concludes the paper.

\section{Semantics-empowered NOMA: Potentials and Limits}

\begin{figure}[ht]
    \centering
    \includegraphics[width=5.5in]{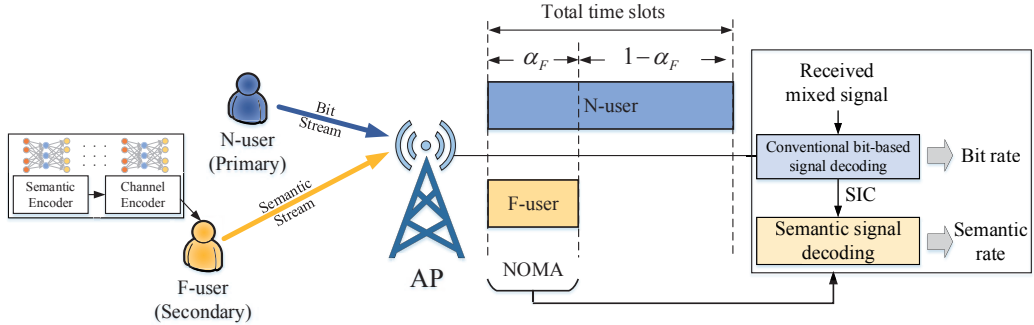}
    \caption{The illustration of the proposed semantics-empowered two-user uplink NOMA framework, where the primary N-user employs BitCom for uploading information and the secondary F-user tries to reuse the N-user's resource block to upload information with the aid of SemCom.}\label{model1}
\end{figure}
\subsection{System Model}
As shown in Fig. \ref{model1}, we propose a novel semantics-empowered two-user uplink NOMA framework, which consists one single-antenna AP and two single-antenna users. In terms of the distance between the AP and the user, we refer to the two users as N-user and F-user. In particular, it is assumed that the N-user is the primary user which is allocated with a dedicated resource block for uploading information to the AP. The F-user is assumed to be a secondary user which aims to reuse the primary N-user's resource block via NOMA. Motivated by the fact that SemCom is usually superior to BitCom in the low SNR regime~\cite{9398576}, in this paper, SemCom is employed at the secondary F-user for improving the transmission efficiency since it has with a limited channel condition given the long distance to the AP. Moreover, the primary N-user having a high channel condition is assumed to employ the BitCom for uploading information as that in conventional NOMA. Let ${h_n}$ and ${h_f}$ denote the channel coefficient from the N-user to the AP and from the F-user to the AP, respectively. To study the maximum performance limit, perfect CSI is assumed in this paper. Let $x_n$ and $x_f$ denote the transmit symbols for BitCom and SemCom, respectively, which are normalized to unit power. Therefore, if the F-user is admitted to the channel via NOMA, the received signal at the AP is given by 
\begin{equation}\label{received}
y = \sqrt{{p_n}}{x_n} + \sqrt{{p_f}}{x_f} + z,
\end{equation}
where ${p_n} \in \left[ {0,P_n^{\max }} \right]$ and ${p_f} \in \left[ {0,P_f^{\max }} \right]$ denote the employed transmit power at the N-user and F-user, respectively. Here, $P_n^{\max }$ and $P_f^{\max }$ represent the corresponding maximum transit power budget. $z$ denotes the additive white Gaussian noise (AWGN) at the AP with mean zero and variance $\sigma^2$.

Following the principle of uplink NOMA, SIC is employed at the AP to successively decode the two users' signals with a specific decoding order. Recall the fact that the N-user using BitCom generally has a higher channel condition than the F-user using SemCom, the conventional channel condition based SIC decoding order is assumed~\cite{Liu2017,Ding}, i.e., the AP first decodes the bit signal uploaded by the N-user and then decodes the semantic signal uploaded by the F-user by removing the bit signal from the received mixed signal with the aid of SIC. It can be observed that using the aforementioned SIC decoding order, the primary N-user will suffer from additional interference caused by admitting the secondary F-user user. Therefore, for better protecting the communication performance of the primary N-user, as shown in Fig. \ref{model1}, the secondary F-user can be admitted using part of N-user's entire time slots. Here, the length of the entire time slots is equal to one specific channel block time. For simplicity, we assume a normalized unit channel block time, i.e., $T=1$. Let $0 \le \alpha_f \le 1$ denote the time portion allocated to the F-user. As a result, the entire time slots can be divided into non-orthogonal time slots and N-user-only time slots. Based on the proposed framework, in the following, we discuss the bit rate and semantic rate achieved at the N-user and F-user, respectively.

\subsubsection{Achieved bit rate at the N-user} As previously described, during the non-orthogonal time slots, the AP decodes the N-user's bit signal under the interference of the F-user's semantic signal. As conventional BitCom is employed at the N-user, following the Shannon classical information theory, the achieved bit rate (bits/s/Hz) over the time portion $\alpha_f$ is given by
\begin{equation}\label{r_NOMA1}
R_1 = {\alpha _f}{\log _2}\left( {1 + \frac{{{p_n}{{\left| {{h_n}} \right|}^2}}}{{{p_f}{{\left| {{h_f}} \right|}^2} + {\sigma ^2}}}} \right).
\end{equation}
Next, during the remaining N-user-only time slots, only the N-user uploads its bit stream to the AP and the F-user is not allowed to transmit. The corresponding achieved bit rate over the time portion $(1 - \alpha_f )$ is given by 
\begin{equation}\label{r_NOMA2}
R_2 = {\left(1 - \alpha_f \right)}{\log _2}\left( {1 + \frac{{{p_n}{{\left| {{h_n}} \right|}^2}}}{{{\sigma ^2}}}} \right).
\end{equation}
As a result, the overall achieved bit rate of the N-user over the entire time slots is calculated as follows:
\begin{equation}\label{r_NOMA}
R ={\alpha _f}{\log _2}\left( {1 + \frac{{{p_n}{{\left| {{h_n}} \right|}^2}}}{{{p_f}{{\left| {{h_f}} \right|}^2} + {\sigma ^2}}}} \right)+{\left(1 - \alpha_f \right)}{\log _2}\left( {1 + \frac{{{p_n}{{\left| {{h_n}} \right|}^2}}}{{{\sigma ^2}}}} \right).
\end{equation}

\subsubsection{Achieved semantic rate at the F-user} For the employed SemCom at the F-user, we assumed that the text transmission is carried out by using the DeepSC proposed in~\cite{9398576}. For characterizing the performance of DeepSC semantic text transmission, the semantic rate (susts/s/Hz) over the time portion $\alpha_f$ is defined as follows~\cite{9763856}:
\begin{equation}
{\overline S}=\frac{{\alpha_f}I}{KL} \varepsilon \left( K, \gamma \right),
\end{equation}
where $L$ denotes the average number of words per sentence to be transmitted, $K \in \mathbb{Z}^{+}$ denotes the average number of mapped semantic symbols to be transmitted for each word by DeepSC, $I$ (semantic units (suts)) denotes the average amount of semantic information contained in the transmitted sentence, and $\varepsilon_K(\gamma)$ denotes the semantic similarity function with respect to $K$ and the received SNR, $\gamma=\frac{{p_f}{\left\lvert {h_f} \right\rvert}^2}{\sigma^2}$. In general, the value of $\varepsilon (K,\gamma)$ under the given $K$ and $\gamma$ can only be obtained by running the DeepSC tool under the considered channel condition. To address the difficulty of lacking a closed-form expression, the data regression method was proposed in \cite{Mu}. For a given $K$, $\varepsilon (K,\gamma)$ is approximated by a generalized logistic function as follows~\cite{Mu}:
\begin{align}\label{logistic}
\varepsilon \left( K,\gamma \right) \approx {\widetilde \varepsilon _K}\left( \gamma  \right) \triangleq {A_{K,1}} + \frac{{{A_{K,2}} - {A_{K,1}}}}{{1 + {e^{ - \left( {{C_{K,1}}\gamma  + {C_{K,2}}} \right)}}}},
\end{align}
where ${A_{K,1}}>0$ and ${A_{K,2}}>0$ denote the lower (left) asymptote and the upper (right) asymptote, respectively, ${{C_{K,1}}}>0$ denotes the logistic growth rate, and ${{C_{K,2}}}$ controls the logistic mid-point for different $K$. As presented in \cite{Mu}, the approximation can achieve high accuracy for each $K$ and provides a tractable form. Therefore, the achieved semantic rate at the F-user can be approximated by ${\overline S} \approx \frac{{\alpha_f}I}{KL} {\widetilde \varepsilon _K} \left(\frac{{p_f}{\left\lvert {h_f} \right\rvert}^2}{\sigma^2}\right)$. Note that in practice, the achieved semantic similarity should be individually considered for ensuring the valid performance of SemCom. Therefore, in this paper, we define the following \emph{effective semantic rate}\footnote{In the following context, we use ``semantic rate'' for referring to ``effective semantic rate'' for convenience.}:
\begin{equation}\label{s_NOMA}
S=\frac{{\alpha_f}I}{KL} {\widetilde \varepsilon _K} \left(\frac{{p_f}{\left\lvert {h_f} \right\rvert}^2}{\sigma^2}\right){\mathfrak{1}}\left({{{\widetilde \varepsilon }_K} \ge \overline \varepsilon }\right),
\end{equation}
where ${\overline \varepsilon }$ denotes the minimum required semantic similarity for achieving valid SemCom. Here, ${\mathfrak{1}}\left( \cdot \right)$ is a binary indicator function, whose value is 1 if ${{{\widetilde \varepsilon }_K} \ge \overline \varepsilon }$; and the value is 0, otherwise. 

\subsubsection{Achieved semantic-versus-bit rate region} To study the fundamental performance limit of the proposed semantics-empowered uplink NOMA framework, the achieved SvB rate region is defined as follows: 
\begin{align}\label{Region_NOMA}
\begin{gathered}
{{\mathcal{R}}}_{{\rm{SvB}}} = \bigcup\limits_{{p_f} \in \left[ {0,P_f^{\max }} \right],{\alpha _f} \in \left[ {0,1} \right]} {\left\{ {\left( {S,R} \right):S \le \frac{{{\alpha _f}I}}{{KL}}{{\widetilde \varepsilon }_K}(\frac{{{p_f}{{\left| {{h_f}} \right|}^2}}}{{{\sigma ^2}}}){\mathfrak{1}}({{\widetilde \varepsilon }_K} \ge \overline \varepsilon ),} \right.}  \hfill \\
\left. {R \le {\alpha _f}{{\log }_2}\left( {1 + \frac{{P_n^{\max }{{\left| {{h_n}} \right|}^2}}}{{{p_f}{{\left| {{h_f}} \right|}^2} + {\sigma ^2}}}} \right) + \left( {1 - {\alpha _f}} \right){{\log }_2}\left( {1 + \frac{{P_n^{\max }{{\left| {{h_n}} \right|}^2}}}{{{\sigma ^2}}}} \right)} \right\}, \hfill \\ 
\end{gathered} 
\end{align}
which consists of all the semantic and bit rate-pairs achieved by the two users over the given resource block. Here, we assume that ${\widetilde \varepsilon _K}\left(\frac{{P_f^{\max }{{\left| {{h_f}} \right|}^2}}}{{{\sigma ^2}}}\right) \ge \overline \varepsilon $ to ensure the feasibility of SemCom at the F-user. To characterize the boundary of ${{\mathcal{R}}}_{{\rm{SvB}}}$, we consider the following optimization problem:
\begin{subequations}\label{NOMA boundary}
\begin{align}
\mathop {\max }\limits_{{p_f} \in \left[ {0,P_f^{\max }} \right],{\alpha _f} \in \left[ {0,1} \right]} &\;\frac{{{\alpha _f}I}}{{KL}}{\widetilde \varepsilon _K}\left(\frac{{{p_f}{{\left| {{h_f}} \right|}^2}}}{{{\sigma ^2}}}\right){\mathfrak{1}}\left( {\widetilde \varepsilon _K} \ge \overline \varepsilon \right)\\
\label{S NOMA 1}{\rm{s.t.}}\;\;&{\alpha _f}{\log _2}\left( {1 + \frac{{P_n^{\max }{{\left| {{h_n}} \right|}^2}}}{{{p_f}{{\left| {{h_f}} \right|}^2} + {\sigma ^2}}}} \right) + \left( {1 - {\alpha _f}} \right){\log _2}\left( {1 + \frac{{P_n^{\max }{{\left| {{h_n}} \right|}^2}}}{{{\sigma ^2}}}} \right) \ge \overline R,
\end{align}
\end{subequations}
where $\overline R$ is a target bit rate required by the N-user. By solving problem \eqref{NOMA boundary} for all $\overline R  \in \left[ {0,{{\log }_2}\left( {1 + \frac{{P_n^{\max }{{\left| {{h_n}} \right|}^2}}}{{{\sigma ^2}}}} \right)} \right]$, the complete boundary of ${{\mathcal{R}}}_{{\rm{SvB}}}^{\rm{N}}$ can be characterized. For any given $\overline R$, problem \eqref{NOMA boundary} can be solved by exhaustively searching over ${p_f} \in \left[ {0,P_f^{\max }} \right]$ and ${\alpha _f} \in \left[ {0,1} \right]$.

\subsection{Performance Comparison}\label{PC}
\subsubsection{SemCom v.s. BitCom} For evaluating the performance gain introduce by SemCom, we further consider the conventional case, where the F-use also employs BitCom for uploading information together with the N-user in NOMA. In this case, the achieved bit rate of the F-user is given by 
\begin{equation}
R^{\rm{B}}={\alpha_f}\log_2\left( 1+\frac{{p_f}\left\lvert {h_f} \right\rvert^2 }{\sigma^2} \right). 
\end{equation}
In order to conduct a fair comparison between SemCom and BitCom, $R^{\rm{B}}$ is transformed into the following equivalent semantic rate~\cite{9763856}:   
\begin{equation}\label{B_SR}
S^{\rm{B}}=R^{\rm{B}}\frac{I}{{\mu}L}{\varepsilon_C},
\end{equation}
where ${\mu}$ (bits/word) denotes the average number of bits to be transmitted per word for the text transmission and $\varepsilon_C$ denotes the semantic similarity achieved by BitCom, which depends on the achieved bit error. It is worth noting that ${\mu} \gg K$ in practice, i.e., for transmitting one word, BitCom requires much larger number of bits to convey than the number of semantic symbols required by SemCom. Therefore, we can compare \eqref{s_NOMA} and \eqref{B_SR} as follows:
\begin{equation}\label{difference}
{\Delta = {S^{\rm{B}}} - {S}} = \left\{ {\frac{1}{\mu }{{\log }_2}\left( {1 + {\gamma _f}} \right){\varepsilon _C} - \frac{1}{K}{{\widetilde \varepsilon }_K}\left( {{\gamma _f}} \right)} \right\}\frac{{{\alpha_f}}I}{L},
\end{equation}
where ${\gamma _f} = \frac{{{p_f}{{\left| {{h_f}} \right|}^2}}}{{{\sigma ^2}}}$ denotes the received SNR and the semantic similarity requirement in \eqref{s_NOMA} is dropped for simplicity. Let $s^{\rm{B}}=\frac{1}{\mu }{{\log }_2}\left( {1 + {\gamma _f}} \right){\varepsilon _C}$ and $s=\frac{1}{K}{{\widetilde \varepsilon }_K}\left( {{\gamma _f}} \right)$. The individual advantages of SemCom and BitCom can be demonstrated by considering the follow two cases. First, we consider an extreme case ${\gamma _f} \to \infty $, we have ${s_\infty^{\rm{B}}}\to \infty$ and ${s_\infty}\to \frac{1}{K}$, which means that $\Delta_\infty  \to \infty $, i.e., BitCom outperforms SemCom. Second, we consider the case of ${\gamma _f}=0$ dB. According to the results presented in \cite{Mu}, ${{\widetilde \varepsilon }_K}\left( {{\gamma _f}} \right)=0.5$ when $K=4$, i.e., ${s_0}=\frac{1}{8}$ for SemCom. However, for BitCom, ${s_0^{\rm{B}}} = \frac{{{\varepsilon _C}}}{\mu } \le \frac{1}{\mu }$. For example, assuming that one word contains five letters and ASCII code is used to encode each letter, it results in ${\mu }=40$. As a result, we can conclude that $\Delta_0 <0$, i.e., SemCom outperforms BitCom. Given the above two cases, it shows that SemCom and BitCom are generally preferable in the low- and high-SNR cases, respectively. This is also consistent with the results of existing research contributions on SemCom~\cite{9398576}. 
\begin{figure}[!t]
  \centering
  \includegraphics[width=3.5in]{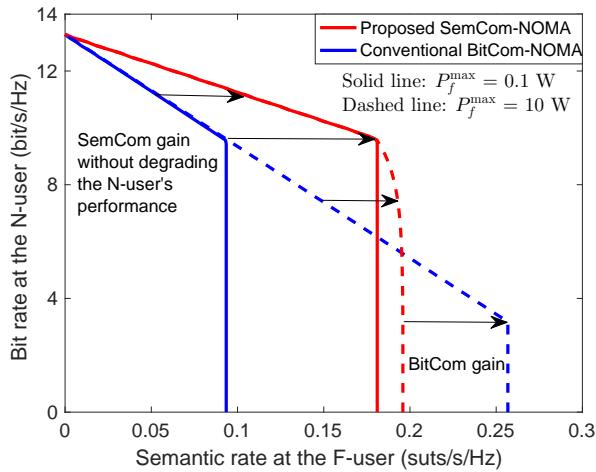}\\
  \caption{SvB rate region comparison of the proposed semantics-empowered NOMA and conventional BitCom-based NOMA.}\label{region_C}
\end{figure}
\subsubsection{Numerical Examples} To further illustrate the performance comparison between SemCom and BitCom, we present numerical examples in Fig. \ref{region_C} for comparing the SvB rate region achieved by the proposed semantics-empowered NOMA and the equivalent SvB rate region achieved by the conventional BitCom-based NOMA. The equivalent SvB rate region can be characterized by replacing the objective function of problem \eqref{NOMA boundary} with the equivalent semantic rate \eqref{B_SR}. For the employed SemCom, we set ${\overline \varepsilon }=0.9$ and $K=5$. For BitCom, we set ${\varepsilon_C}=1$, i.e., there is no bit error, and ${\mu }=40$~\cite{9763856}. It is assumed that the distances from the AP to the N-user and F-user are 10 meters and 30 meters, respectively. The distance-dependent path loss is modelled as $\rho  = {\rho _0}{\left( {{1 \mathord{\left/
{\vphantom {1 d}} \right.
\kern-\nulldelimiterspace} d}} \right)^\beta }$, where ${\rho _0}=-30$ dB denotes the reference path loss at 1 meter, $\beta =4$ denotes the path loss exponent, and $d$ denotes the link distance in meters. The noise power is set to $\sigma^2 = - 80$ dBm. We fix $P_n^{\max}=1$ W and consider two cases of $P_f^{\max}=0.1$ W and $P_f^{\max}=10$ W. As can be observed from Fig. \ref{region_C}, it can be observed that when $P_f^{\max}=0.1$ W, the SvB achieved by the proposed semantics-empowered NOMA strictly contains that achieved by the conventional BitCom-based NOMA. In particular, given the same bit rate achieved at the N-user, SemCom enables the F-user to achieve significantly higher performance than BitCom. The reason behind this can be explained as follows. On the one hand, although the limited transmit power (i.e., $P_f^{\max}=0.1$ W) makes the F-user have a high probability to be admitted to the channel, it strictly limits the performance of the F-user if using BitCom. Nevertheless, SemCom comes to the rescue by enabling the F-user to achieve higher communication performance using the limited transmit power. On the other hand, when the primary N-user has a high communication requirement, the permitted transmit power for the secondary F-user will be strictly capped for not causing too much interference to the N-user if admitted. Compared to BitCom, SemCom also allows the F-user to achieve higher communication performance with the small permitted transmit power. As a result, it can be also observed that even when $P_f^{\max}=10$ W, SemCom still significantly outperform BitCom when the N-user communication requirement is high. However, when the N-user only has a small communication requirement (i.e., the permitted transmit power for the F-user is high), it can be observed that BitCom outperforms SemCom whose maximum performance is upper bounded, which is consistent with the above theoretical performance comparison between SemCom and BitCom. The presented results clearly show the attractive benefits and also the potential limits of employing SemCom in NOMA.

\section{An Opportunistic SemCom and BitCom Scheme over Fading Channels} 
\begin{figure}[!h]
  \centering
  \includegraphics[width=5in]{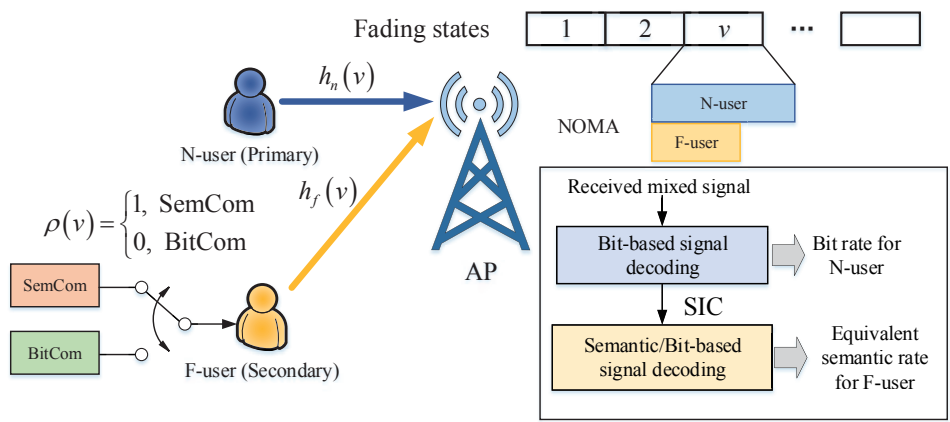}
  \caption{The semantics-empowered two-user uplink NOMA over the fading channel, where the secondary F-user employs an opportunistic SemCom and BitCom scheme for each fading state.}\label{model2}
\end{figure}
In this section, we investigate the proposed semantics-empowered NOMA transmission over fading channels. The channel from the N/F-user to the AP is assumed to follow the quasi-static block fading channel model. The channel remains constant for each fading state $v$ and independently varies from block to block. For any fading state $v$, let ${h_n}\left( v \right)$ and ${h_f}\left( v \right)$ denote the instantaneous channel coefficient of the N-user-AP link and the F-user-AP link, respectively. Here, we assume that the primary N-user communications with the AP via BitCom with a constant transmit power, $P_n$, and the secondary F-user opportunistically reuses the N-user's resource block for uploading information. For managing the interference caused to the N-user in NOMA and ensuring its own achieved communication performance, the F-user should employ appropriate \emph{resource management} (i.e., power control and time scheduling) and select the communication method from SemCom and BitCom. Let ${p}\left( v \right)$ and ${\alpha}\left( v \right)$ denote the employed transmit power and occupied transmission time portion of the F-user at the fading state $v$.
\subsection{Communication Model}
\subsubsection{BitCom at the N-user} As BitCom is employed at the N-user, the instantaneous bit rate achieved at fading state $v$ is given by
\begin{equation}\label{B_R_N}
R\left( v \right) = \alpha \left( v \right){\log _2}\left( {1 + \frac{{{P_n}{{\left| {{h_n}\left( v \right)} \right|}^2}}}{{p\left( v \right){{\left| {{h_f}\left( v \right)} \right|}^2} + {\sigma ^2}}}} \right) + \left( {1 - \alpha \left( v \right)} \right){\log _2}\left( {1 + \frac{{{P_n}{{\left| {{h_n}\left( v \right)} \right|}^2}}}{{{\sigma ^2}}}} \right).
\end{equation}
Accordingly, the ergodic bit rate of the N-user over all the fading states is given by ${{\mathbb{E}}_v}\left[ {R\left( v \right) } \right]$, where ${{\mathbb{E}}_v}\left[  \cdot  \right]$ denotes the expectation over $v$.
\subsubsection{Opportunistic SemCom and BitCom at the F-user} Motivated by the revealed potentials and limits of SemCom in Section \ref{PC}, we propose an opportunistic SemCom and BitCom scheme for the F-user to fully reap the benefits of the two communication methods at each fading state. Let $\rho  \left( v \right)$ denote an indicator function to specify the employed communication method at the F-user, which is defined as follows:
\begin{equation}
  \rho \left( v \right) = \left\{ \begin{gathered}
  1,\;\;{\rm{SemCom}}\;{\rm{is}}\;{\rm{used}} \hfill \\
  0,\;\;{\rm{BitCom}}\;{\rm{is}}\;{\rm{used}} \hfill \\ 
\end{gathered}  \right..
\end{equation}
After removing the N-user's bit signal with the aid of SIC, the AP can decode the F-user's signal in an intereference-free manner at each fading state. By employing the semantic rate defined for SemCom and the equivalent semantic rate defined for BitCom in the previous section, the instantaneous semantic rate of the F-user achievable at fading state $v$ is given by 
\begin{equation}\label{S_R_N}
S\left( v \right) = \alpha \left( v \right)\left\{ {\rho \left( v \right){S_s}\left( v \right) + \left( {1 - \rho \left( v \right)} \right){S_b}\left( v \right)} \right\},
\end{equation}
where ${S_s}\left( v \right) = \frac{I}{{KL}}{\widetilde \varepsilon _K}\left( {\frac{{p\left( v \right){{\left| {{h_f}\left( v \right)} \right|}^2}}}{{{\sigma ^2}}}} \right){\mathfrak{1}}\left( {{{\widetilde \varepsilon }_K} \ge \overline \varepsilon } \right)$ and ${S_b}\left( v \right) = \frac{I}{{\mu L}}{\log _2}\left( {1 + \frac{{p\left( v \right){{\left| {{h_f}\left( v \right)} \right|}^2}}}{{{\sigma ^2}}}} \right)$.  Here, we assume that there is no bit error in the BitCom. Therefore, the ergodic (equivalent) semantic rate of the F-user over all the fading states is given by ${{\mathbb{E}}_v}\left[ {S\left( v \right) } \right]$. 

For the proposed opportunistic SemCom and BitCom scheme, we further consider the following two resource management scenarios:
\begin{figure}[ht]
  \centering
  \includegraphics[width=5.5in]{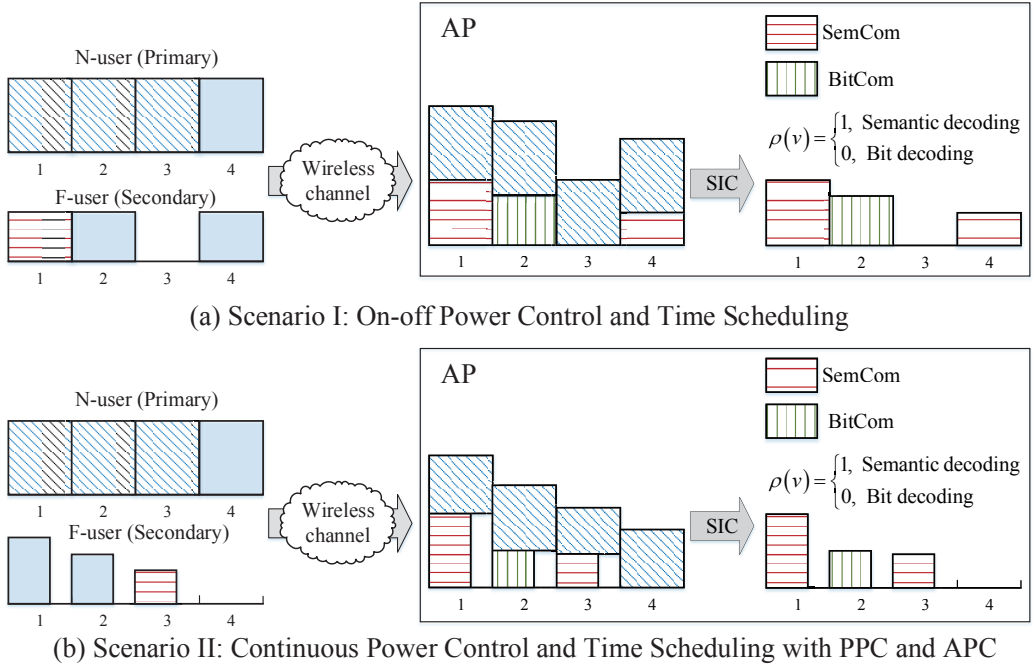}
  \caption{The proposed opportunistic SemCom and BitCom scheme in two scenarios.}\label{protocol}
\end{figure}
\begin{itemize}
  \item \textbf{Scenario I (On-off Power Control and Time Scheduling)}: In this case, the F-user reuses the N-user's resource block via NOMA following an on-off manner. Specifically, the feasible set of the time scheduling and power control is given by 
  \begin{equation}\label{F1}
  {{\mathcal{F}}_1} = \left\{ {\alpha \left( v \right) \in \left\{ {0,1} \right\},p\left( v \right) \in \left\{ {0,{P_0}} \right\},\forall v} \right\},
  \end{equation}
  which means if the F-user wants to reuse N-user's resource block to upload information at fading state $v$, it will occupy the entire transmission time and use a constant transmit power, $P_0$; Otherwise, the F-user remains in silence. Fig. \ref{protocol}(a) illustrates an example of the on-off power control and time scheduling in the proposed opportunistic SemCom and BitCom scheme. The F-user employs SemCom and BitCom at fading states 1\&4 and 2, respectively, and turns off at fading state 3. As long as the F-user is active, it will use a constant transmit power and occupy the entire transmission time. In this scenario, the resource management at the F-user is easy to implement, but reduces the design flexibility.
  \item  \textbf{Scenario II (Continuous Power Control and Time Scheduling)}: In this case, the transmit power and time scheduling of the F-user can be continuously adjusted. Specifically, we considered two types of transmit power constraints, namely the peak power constraint (PPC) and the average power constraint (APC). The PPC limits the instantaneous transmit power employed at the F-user in each fading state, i.e., $\left\{ {p\left( v \right) \le \widehat P,\forall v} \right\}$. The APC limits the long-term transmit power employed at the F-user over all fading states, i.e., ${{\mathbb{E}}_v}\left[{\alpha \left( v \right)} {p\left( v \right)} \right] \le \overline P $. Without loss of generality, we assume that $\overline P \le \widehat P$. The corresponding feasible set of the time scheduling and power control is given by
  \begin{equation}\label{F2}
  {{\mathcal{F}}_2} = \left\{ {0 \le \alpha \left( v \right) \le 1,p\left( v \right) \le \widehat P,{{\mathbb{E}}_v}\left[ {\alpha \left( v \right)}{p\left( v \right)} \right] \le \overline P ,\forall v} \right\}.
  \end{equation}
  Fig. \ref{protocol}(b) illustrates an example of the continuous power control and time scheduling. The F-user uses SemCom and BitCom with different transmit power and transmission time at fading states 1\&3 and 2, respectively, and remains in silence at fading state 4. In this scenario, the resource management at the F-user can be design with a high flexibility is easy to implement, but leads to a high complexity for implementation.  
\end{itemize} 
\subsection{Problem Formulation}
For each scenario, our goal is to maximize the ergodic semantic rate of the F-user, $\overline S$, by jointly optimizing the communication method, $\left\{ {\rho \left( v \right),\forall v} \right\}$, power control, $\left\{ {p\left( v \right),\forall v} \right\}$, and time scheduling, $\left\{ {\alpha \left( v \right),\forall v} \right\}$, of each fading state, subject to a minimum ergodic rate constraint of the N-user. As a result, the optimization problem can be formulated as follows:
\begin{subequations}\label{Problem}
\begin{align}
&\mathop {\max }\limits_{\left\{ {\rho \left( v \right),\alpha \left( v \right),p\left( v \right)} \right\}} \;{{\mathbb{E}}_v}\left[ {S\left( v \right)} \right]\\
\label{C1}{\rm{s.t.}}\;\;&{{\mathbb{E}}_v}\left[ {R\left( v \right)} \right] \ge \overline R ,\\
\label{C2}&{\rho\left( v \right) \in \left\{ {0,1} \right\},\forall v}, \\
\label{C3}&\left\{ {\alpha \left( v \right),p\left( v \right)} \right\} \in {{\mathcal{F}}_{\rm{X}}},
\end{align}
\end{subequations}
where ${\rm{X}} \in \left\{ {{\rm{I}},{\rm{II}}} \right\}$ specifies Scenario I or II. It can be observed that problem \eqref{Problem} in each scenario is non-convex given the non-concave objective function, non-convex constraint \eqref{C1}, and the integer optimization variables. However, it can be verified that problem \eqref{Problem} satisfies the ``time-sharing'' condition \cite{timeshare}. To demonstrate this, we take problem under Scenario II as an example. Let ${{\overline S}_{{\rm{II}}}}\left( {\overline R ,\overline P } \right)$ denote the optimal objective function's value given the ergodic bit rate constraint and the average transmit power constraint pairs $\left( {\overline R ,\overline P } \right)$, and $\left\{ {{\rho _x}\left( v \right),{\alpha _x}\left( v \right),{p_x}\left( v \right)} \right\}$ and $\left\{ {{\rho _y}\left( v \right),{\alpha _y}\left( v \right),{p_y}\left( v \right)} \right\}$ denote the optimal solutions to problem \eqref{Problem} under constraint pairs $\left( {\overline R_x ,\overline P_x } \right)$ and $\left( {\overline R_y ,\overline P_y } \right)$, respectively. For any $0 \le \theta  \le 1$, it can be verified that there are always exist feasible solutions $\left\{ {{\rho _z}\left( v \right),{\alpha _z}\left( v \right),{p_z}\left( v \right)} \right\}$ such that ${{\mathbb{E}}_v}\left[ {{S_z}\left( v \right)} \right] \ge \theta {{\overline S}_{{\rm{II}}}}\left( {{{\overline R }_x},{{\overline P }_x}} \right) + \left( {1 - \theta } \right){{\overline S}_{{\rm{II}}}}\left( {{{\overline R }_y},{{\overline P }_y}} \right)$, ${{\mathbb{E}}_v}\left[ {{R_z}\left( v \right)} \right] \ge \theta {\overline R _x} + \left( {1 - \theta } \right){\overline R _y}$, and ${{\mathbb{E}}_v}\left[ {\alpha _z}\left( v \right){{p_z}\left( v \right)} \right] \le \theta {\overline P _x} + \left( {1 - \theta } \right){\overline P _y}$, where ${{R_z}\left( v \right)}$ and ${{S_z}\left( v \right)}$ are defined in \eqref{B_R_N} and \eqref{S_R_N}, respectively. According to the convex analysis in \cite{timeshare}, if one optimization problem satisfies the ``time-sharing'' condition, the duality gap between the primal problem and its Lagrange dual problem is zero, i.e., strong duality holds~\cite{convex}, regardless of the convexity of the original problem. Therefore, we can optimally solve problem \eqref{Problem} via its dual problem.
\section{Ergodic Semantic Rate Maximization for Scenarios I and II}
In this section, we employ the Lagrange duality method to solve problem \eqref{Problem} in Scenarios I and II to derive the optimal communication policy at the F-user for maximizing the achieved ergodic semantic rate.
\subsection{Optimal Solution in Scenario I}
For Scenario I, problem \eqref{Problem} can be simplified into 
\begin{subequations}\label{Problem1}
\begin{align}
&\mathop {\max }\limits_{\left\{ {\rho \left( v \right),p\left( v \right)} \right\}} \;{{\mathbb{E}}_v}\left[ {{\rho \left( v \right){S_s}\left( v \right) + \left( {1 - \rho \left( v \right)} \right){S_b}\left( v \right)}} \right]\\
\label{C11}{\rm{s.t.}}\;\;&{{\mathbb{E}}_v}\left[ {\widetilde R\left( v \right)} \right] \ge \overline R ,\\
\label{C21}&{\rho\left( v \right) \in \left\{ {0,1} \right\},\forall v}, \\
\label{C31}&p\left( v \right) \in \left\{ {0,{P_0}} \right\},\forall v,
\end{align}
\end{subequations}
where $\widetilde R\left( v \right)={\log _2}\left( {1 + \frac{{{P_n}{{\left| {{h_n}\left( v \right)} \right|}^2}}}{{p\left( v \right){{\left| {{h_f}\left( v \right)} \right|}^2} + {\sigma ^2}}}} \right)$. Here, the on-off time scheduling variables $\left\{ {\alpha \left( v \right)} \right\}$ can be safely removed. This is because the case of ${\alpha \left( v \right) = 0}$ is equivalent to ${p \left( v \right) = 0}$ where ${S_s}\left( v \right) = {S_b}\left( v \right) = 0$ and the case of ${\alpha \left( v \right) = 1}$ is equivalent to ${p \left( v \right) = {P_0}}$, i.e., $\left\{ {\alpha \left( v \right)} \right\}$ become dummy variables. 

By applying the Lagrangian dual method, the Lagrangian of problem \eqref{Problem1} is given by
\begin{equation}\label{Lagrangian1}
{{\mathcal{L}}_1}\left( {\rho \left( v \right),p\left( v \right),\lambda } \right) = {{\mathbb{E}}_v}\left[ {\rho \left( v \right){S_s}\left( v \right) + \left( {1 - \rho \left( v \right)} \right){{S}_b}\left( v \right)} \right] + \lambda \left\{ {{{\mathbb{E}}_v}\left[ {\widetilde R\left( v \right)} \right] - \overline R} \right\},
\end{equation}
where $\lambda$ is the non-negative Lagrange multiplier associated with the ergodic bit rate constraint \eqref{C11}. Then, the Lagrange dual function of problem \eqref{Problem1} is expressed as
\begin{equation}\label{Lagrangian_dual1}
g_1\left( \lambda  \right) = \mathop {\max }\limits_{\rho \left( v \right) \in \left\{ {0,1} \right\},p\left( v \right) \in \left\{ {0,{P_0}} \right\},\forall v} {\mathcal{L}}_1\left( {\rho \left( v \right),p\left( v \right),\lambda } \right).
\end{equation}
The maximization of problem \eqref{Lagrangian_dual1} can be decomposed into a number of subproblems having the same structure and each of them corresponds to one fading state. Therefore, the fading index $v$ can be safely dropped for ease of exposition. Focusing on one particular fading state, the associated subproblem for any given $\lambda$ is given by
\begin{equation}\label{Lagrangian_dual1_sub}
\mathop {\max }\limits_{\rho  \in \left\{ {0,1} \right\},p \in \left\{ {0,{P_0}} \right\}} {\overline {\mathcal{L}} _1}\left( {\rho ,p} \right),
\end{equation}
where ${\overline {\mathcal{L}} _1}\left( {\rho ,p} \right) = \rho {S_s} + \left( {1 - \rho } \right){S_b} + \lambda \widetilde R$. To solve problem \eqref{Lagrangian_dual1_sub}, we need to compare the values of ${\overline {\mathcal{L}} _1}\left( {\rho ,p} \right)$ achieved by $\left( {\rho  = 1,p = {P_0}} \right)$, $\left( {\rho  = 0,p = {P_0}} \right)$, $\left( {\rho  = 1,p =0} \right)$ and $\left( {\rho  = 0,p =0} \right)$. It is worth noting that the value of ${\overline {\mathcal{L}} _1}\left( {\rho ,p} \right)$ remains unchanged for $\left( {\rho  = 1,p =0} \right)$ and $\left( {\rho  = 0,p =0} \right)$, we ignore the case of $\left( {\rho  = 1,p =0} \right)$ for ease of exposition. 

When $\rho  = 1$, we have
\begin{equation}\label{p1}
{\overline {\mathcal{L}} _1}\left( {1,{P_0}} \right) = \frac{I}{{KL}}{\widetilde \varepsilon _K}\left( {\frac{{{P_0}{{\left| {{h_f}} \right|}^2}}}{{{\sigma ^2}}}} \right){\mathfrak{1}}\left( {{{\widetilde \varepsilon }_K} \ge \overline \varepsilon } \right) + \lambda {\log _2}\left( {1 + \frac{{{P_n}{{\left| {{h_n}} \right|}^2}}}{{{P_0}{{\left| {{h_f}} \right|}^2} + {\sigma ^2}}}} \right).
\end{equation}

When $\rho  = 0$, we have
\begin{equation}\label{p01}
{\overline {\mathcal{L}} _1}\left( {0,{P_0}} \right) = \frac{I}{{\mu L}}{\log _2}\left( {1 + \frac{{{P_0}{{\left| {{h_f}} \right|}^2}}}{{{\sigma ^2}}}} \right) + \lambda {\log _2}\left( {1 + \frac{{{P_n}{{\left| {{h_n}} \right|}^2}}}{{{P_0}{{\left| {{h_f}} \right|}^2} + {\sigma ^2}}}} \right),
\end{equation}
\begin{equation}\label{p00}
{\overline {\mathcal{L}} _1}\left( {0,0} \right) = \lambda {\log _2}\left( {1 + \frac{{{P_n}{{\left| {{h_n}} \right|}^2}}}{{{\sigma ^2}}}} \right).
\end{equation}
Therefore, the optimal power control for the case of $\rho  = 0$ can be expressed as
\begin{equation}\label{p}
{p_{\rho  = 0}} = \left\{ \begin{gathered}
  {P_0},\;{\rm{if}}\;{\overline {\mathcal{L}} _1}\left( {0,{P_0}} \right) > {\overline {\mathcal{L}} _1}\left( {0,0} \right) \hfill \\
  0,\;{\rm{otherwise}} \hfill \\ 
\end{gathered}  \right..
\end{equation}
Then, the optimal solution to problem \eqref{Lagrangian_dual1_sub} for any fading state $v$ is given by
\begin{subequations}\label{poo}
\begin{align}\label{po1}
{\rho ^*} &= \left\{ \begin{gathered}
  1,\;{\rm{if}}\;{\overline {\mathcal{L}} _1}\left( {1,{P_0}} \right) > {\overline {\mathcal{L}} _1}\left( {0,{p_{\rho  = 0}}} \right) \hfill \\
  0,\;{\rm{otherwise}} \hfill \\ 
\end{gathered}  \right.,\\
\label{po2}
{p^*} &= \left\{ \begin{gathered}
  {P_0},\;{\rm{if}}\;{\rho ^*} = 1\; \hfill \\
  {p_{\rho  = 0}},\;{\rm{otherwise}} \hfill \\ 
\end{gathered}  \right..
\end{align}
\end{subequations}
For any given $\lambda$, we can solve problem \eqref{Lagrangian_dual1} by solving problem \eqref{Lagrangian_dual1_sub} using the results of \eqref{poo} for all fading states. Based on this, we can optimally solve problem \eqref{Problem1} by iteratively solving \eqref{Lagrangian_dual1} under a fixed $\lambda$, and employing the bisection method to update until the ergodic bit rate constraint \eqref{C11} is eventually met with equality.
\subsection{Optimal Solution in Scenario II}
For Scenario II, problem \eqref{Problem} can be expressed as follows: 
\begin{subequations}\label{Problem2}
\begin{align}
&\mathop {\max }\limits_{\left\{ {\rho \left( v \right),{{\alpha }\left( v \right)},p\left( v \right)} \right\}} \;{{\mathbb{E}}_v}\left[ {\alpha \left( v \right)\left\{ {\rho \left( v \right){S_s}\left( v \right) + \left( {1 - \rho \left( v \right)} \right){S_b}\left( v \right)} \right\}} \right]\\
\label{C12}{\rm{s.t.}}\;\;&{{\mathbb{E}}_v}\left[ {R\left( v \right)} \right] \ge \overline R ,\\
\label{C22}&{{\mathbb{E}}_v}\left[ {{\alpha }\left( v \right)}{p\left( v \right)} \right] \le \overline P ,\\
\label{C32}&{\rho\left( v \right) \in \left\{ {0,1} \right\}}, 0 \le \alpha \left( v \right) \le 1, 0 \le p\left( v \right) \le \widehat P, \forall v.
\end{align}
\end{subequations}

By applying the Lagrangian dual method, the Lagrangian of problem \eqref{Problem2} can be expressed as
\begin{equation}\label{Lagrangian2}
\begin{gathered}
  {{{\mathcal{L}}}_2}\left( {\rho \left( v \right),\alpha \left( v \right),p\left( v \right),\lambda } \right) \hfill \\
  \;\;\; = {{\mathbb{E}}_v}\left[ {\alpha \left( v \right)\left\{ {\rho \left( v \right){S_s}\left( v \right) + \left( {1 - \rho \left( v \right)} \right){{S}_b}\left( v \right)} \right\}} \right]\; \hfill \\
  \;\;\;\;\;\;\; + \beta \left\{ {{{\mathbb{E}}_v}\left[ {R\left( v \right)} \right] - \overline R} \right\} + \delta \left\{ {\overline P  - {{\mathbb{E}}_v}\left[{{\alpha }\left( v \right)} {p\left( v \right)} \right]} \right\}, \hfill \\ 
\end{gathered}
\end{equation}
where $\beta$ and $\delta$ are the non-negative Lagrange multipliers associated with the ergodic bit rate constraint \eqref{C12} and the average transmit power constraint \eqref{C22}, respectively. Then, the Lagrange dual function of problem \eqref{Problem2} is expressed as
\begin{equation}\label{Lagrangian_dual2}
{g_2}\left( {\beta ,\delta } \right) = \mathop {\max }\limits_{\rho \left( v \right) \in \left\{ {0,1} \right\},0 \le \alpha \left( v \right) \le 1,0 \le p\left( v \right) \le \widehat P,\forall v} {{{\mathcal{L}}}_2}\left( {\rho \left( v \right),\alpha \left( v \right),p\left( v \right),\beta ,\delta } \right).
\end{equation}
Accordingly, the dual problem of \eqref{Problem2} can be expressed as
\begin{equation}\label{dual2}
\mathop {\min }\limits_{\beta  \ge 0,\delta  \ge 0} \;{g_2}\left( {\beta ,\delta } \right).
\end{equation}

\subsubsection{Obtaining ${g_2}\left( {\beta ,\delta } \right)$ via Solving Problem \eqref{Lagrangian_dual2}} For given dual variables $\beta$ and $\delta$, ${g_2}\left( {\beta ,\delta } \right)$ can be obtained by maximizing the Lagrangian given in \eqref{Lagrangian2}. Similarly, problem \eqref{Lagrangian_dual2} can be decomposed into a number of subproblems, each of which is associated with one particular fading state and shares the same structure. For one particular fading state, the associated subproblem given $\beta$ and $\delta$ is given by
\begin{equation}\label{sub2}
\mathop {\max }\limits_{\rho  \in \left\{ {0,1} \right\},0 \le \alpha  \le 1,0 \le p \le \hat P} {\overline {\mathcal{L}} _2}\left( {\rho ,\alpha ,p} \right),
\end{equation}
where ${\overline {{\mathcal{L}}} _2}\left( {\rho ,\alpha ,p} \right) = \alpha \left\{ {\rho {S_s} + \left( {1 - \rho } \right){S_b}} \right\} + \beta R - \delta \alpha p$. Here, the fading state index $v$ is dropped for brevity. In order to solve problem \eqref{sub2}, we need to compare the optimal values of ${\overline {{\mathcal{L}}} _2}\left( {\rho ,\alpha ,p} \right)$ when $\rho = 1$ and $\rho = 0$. To obtain the corresponding optimal value to problem \eqref{sub2}, we have the following two subproblems. 
\begin{equation}\label{U1}
\begin{gathered}
  {\overline {{\mathcal{L}}} _2^{\rho  = 1}}\left( {\alpha_s^* ,p_s^*} \right) \hfill \\
   = \mathop {\max }\limits_{0 \le \alpha  \le 1,0 \le p \le \hat P} \left\{ {{S_s} + \beta \left\{ {{{\log }_2}\left( {1 + \frac{{{P_n}{{\left| {{h_n}} \right|}^2}}}{{p{{\left| {{h_f}} \right|}^2} + {\sigma ^2}}}} \right) - {{\log }_2}\left( {1 + \frac{{{P_n}{{\left| {{h_n}} \right|}^2}}}{{{\sigma ^2}}}} \right)} \right\} - \delta p} \right\}\alpha,  \hfill \\ 
\end{gathered} 
\end{equation}
\begin{equation}\label{U2}
\begin{gathered}
  {\overline {{\mathcal{L}}} _2^{\rho  = 0}}\left( {\alpha_b^* ,p_b^*} \right) =  \hfill \\
  \mathop {\max }\limits_{0 \le \alpha  \le 1,0 \le p \le \hat P} \left\{ {{S_b} + \beta \left\{ {{{\log }_2}\left( {1 + \frac{{{P_n}{{\left| {{h_n}} \right|}^2}}}{{p{{\left| {{h_f}} \right|}^2} + {\sigma ^2}}}} \right) - {{\log }_2}\left( {1 + \frac{{{P_n}{{\left| {{h_n}} \right|}^2}}}{{{\sigma ^2}}}} \right)} \right\} - \delta p} \right\}\alpha.  \hfill \\ 
\end{gathered} 
\end{equation}
In \eqref{U1} and \eqref{U2}, we ignore the constant term which is not related to $\alpha$ and $p$ for ease of exposition.

Next, we first focus on the case of $\rho = 1$ and address problem \eqref{U1}. Define ${\Pi _s}\left( p \right) \triangleq {S_s} + \beta \left\{ {{{\log }_2}\left( {1 + \frac{{{P_n}{{\left| {{h_n}} \right|}^2}}}{{p{{\left| {{h_f}} \right|}^2} + {\sigma ^2}}}} \right) - {{\log }_2}\left( {1 + \frac{{{P_n}{{\left| {{h_n}} \right|}^2}}}{{{\sigma ^2}}}} \right)} \right\} - \delta p$, which is a function only with respect to $p$. As variable $\alpha$ is non-negative, problem \eqref{U1} can be rewritten as follows:
\begin{equation}\label{U11}
{\overline {{\mathcal{L}}} _2^{\rho  = 1}}\left( {\alpha_s^* ,p_s^*} \right) = \mathop {\max }\limits_{0 \le \alpha  \le 1} \left\{ {\mathop {\max }\limits_{0 \le p \le \hat P} \;{\Pi _s}\left( p \right)} \right\}\alpha. 
\end{equation}
For the maximization of ${{\Pi _s}\left( p \right)}$, the optimal transmit power ${p_s^*}$ can be obtained by applying one-dimensional search. Then, problem \eqref{U11} becomes
\begin{equation}\label{U111}
\mathop {\max }\limits_{0 \le \alpha  \le 1} \;{\Pi _s}\left( {p_s^*} \right)\alpha,
\end{equation}
It can be observed that problem \eqref{U111} is a linear program (LP) with respect to $\alpha$. Thus, the optimal time scheduling, denoted by $\alpha _s^*$, can be obtained as follows:
\begin{equation}\label{as}
\alpha _s^* = \left\{ \begin{gathered}
  1,\;{\rm{if}}\;{\Pi _s}\left( {p_s^*} \right) > 0, \hfill \\
  {a_0},\;{\rm{if}}\;{\Pi _s}\left( {p_s^*} \right) = 0, \hfill \\
  0,\;{\rm{otherwise}}, \hfill \\ 
\end{gathered}  \right.
\end{equation}
where ${a_0}$ can be any feasible value ranging from 0 to 1 since it does not affect the optimal value of problem \eqref{U111} when ${\Pi _s}\left( {p_s^*} \right) = 0$. In this paper, we set ${a_0}=0$ for simplicity. As a result, the optimal value and corresponding optimal solutions to problem \eqref{U1} are obtained.

Then, we focus on the case of $\rho = 0$ and address problem \eqref{U2}. Similarly, let us define ${\Pi _b}\left( p \right) \triangleq {S_b} + \beta \left\{ {{{\log }_2}\left( {1 + \frac{{{P_n}{{\left| {{h_n}} \right|}^2}}}{{p{{\left| {{h_f}} \right|}^2} + {\sigma ^2}}}} \right) - {{\log }_2}\left( {1 + \frac{{{P_n}{{\left| {{h_n}} \right|}^2}}}{{{\sigma ^2}}}} \right)} \right\} - \delta p$. Problem \eqref{U2} can be solved as follows:
\begin{equation}\label{U22}
{\overline {{\mathcal{L}}} _2^{\rho  = 0}}\left( {\alpha_b^* ,p_b^*} \right) = \mathop {\max }\limits_{0 \le \alpha  \le 1} \left\{ {\mathop {\max }\limits_{0 \le p \le \hat P} \;{\Pi _b}\left( p \right)} \right\}\alpha,
\end{equation}
where the optimal transmit power ${p_b^*}$ can be obtained by applying one-dimensional search for maximizing ${{\Pi _b}\left( p \right)}$. The optimal time scheduling to problem \eqref{U2}, denoted by $\alpha _b^*$, is given by
\begin{equation}\label{ab}
\alpha _b^* = \left\{ \begin{gathered}
  1,\;{\rm{if}}\;{\Pi _b}\left( {p_b^*} \right) > 0, \hfill \\
  {a_0},\;{\rm{if}}\;{\Pi _b}\left( {p_b^*} \right) = 0, \hfill \\
  0,\;{\rm{otherwise}}, \hfill \\ 
\end{gathered}  \right.
\end{equation}

As a result, for any given $\beta$ and $\delta$, the optimal solutions to problem \eqref{sub2} at fading state $v$ are given by
\begin{subequations}\label{roo}
\begin{align}\label{ro}
  &{\rho ^*} = \left\{ \begin{gathered}
  1,\;{\rm{if}}\;\overline {{\mathcal{L}}} _2^{\rho  = 1}\left( {\alpha _s^*,p_s^*} \right) > \overline {{\mathcal{L}}} _2^{\rho  = 0}\left( {\alpha _b^*,p_b^*} \right), \hfill \\
  0,\;{\rm{otherwise}}, \hfill \\ 
\end{gathered}  \right.\\
\label{ap}
&\left( {{a^*},{p^*}} \right) = \left\{ \begin{gathered}
  \left( {\alpha _s^*,p_s^*} \right),\;{\rm{if}}\;{\rho ^*} = 1, \hfill \\
  \left( {\alpha _b^*,p_b^*} \right),\;{\rm{if}}\;{\rho ^*} = 0. \hfill \\ 
\end{gathered}  \right.
\end{align}
\end{subequations}
It is worth mentioning that the time scheduling solution in \eqref{ap} may cannot provide the optimal primal solution to problem \eqref{Problem2} even given the optimal dual variables $\beta^*$ and $\delta^*$. However, by substituting the optimal solutions provided in \eqref{roo} for each fading state into problem \eqref{sub2}, the dual function ${g_2}\left( {\beta ,\delta } \right)$ can be obtained.

\subsubsection{Finding Optimal Dual Solutions $\beta^*$ and $\delta^*$ to Problem \eqref{dual2}} With the obtained optimal solutions $\left\{ {{\rho ^*}\left( v \right),{\alpha ^*}\left( v \right),{p^*}\left( v \right)} \right\}$ for given $\beta$ and $\delta$, we employ sub-gradient based methods, e.g., ellipsoid method~\cite{Ellipsoid}, to iteratively solve the dual problem \eqref{dual2}. In each iteration, the used sub-gradient for updating $\left( {\beta ,\delta } \right)$ is denoted by $\left( {\Delta \beta ,\Delta \delta } \right)$, which is given by
\begin{subequations}\label{sub-gradient}
\begin{align}\label{beta}
&\Delta \beta  = {{\mathbb{E}}_v}\left[ {{R^*}\left( v \right)} \right] - \overline R,\\
\label{de}
&\Delta \delta  = \overline P  - {{\mathbb{E}}_v}\left[ {{{\alpha ^*}\left( v \right)}}{{p^*}\left( v \right)} \right],
\end{align}
\end{subequations}
where ${{R^*}\left( v \right)}$ denotes the achieved bit rate at the fading state $v$ using the obtained solutions $\left( {{\rho ^*}\left( v \right),{\alpha ^*}\left( v \right),{p^*}\left( v \right)} \right)$. Using the above subgradients, the dual variables can be updated by the ellipsoid method towards the optimal solutions, which are denoted by $\beta^*$ and $\delta^*$.  

\subsubsection{Constructing Optimal Primal Solution to Problem \eqref{Problem2}} By exploiting the obtained optimal dual variables $\beta^*$ and $\delta^*$, we continue to find the optimal primal solutions to problem \eqref{Problem2}. Recalling the fact that for a convex optimization problem, the optimal solution which maximizes the Lagrange function under the optimal dual solution is the optimal primal solution if and only if such a solution is unique and primal feasible~\cite{convex}. In our considered problem, the optimal solution $\left\{ {{\alpha ^*}\left( v \right)} \right\}$ may not be unique when ${\Pi _s}\left( {p_s^*} \right) = 0$ or ${\Pi _b}\left( {p_b^*} \right) = 0$, see \eqref{as} and \eqref{ab}. As a result, additional steps are required to construct the optimal primal solution. It can be observed that given optimal dual variables $\beta^*$ and $\delta^*$, the communication method indicator, $\left\{ {{\rho ^*}\left( v \right)} \right\}$, and the transmit power, $\left\{ {{p^*}\left( v \right)} \right\}$, can be uniquely determined by \eqref{roo}. By substituting $\left\{ {{\rho ^*}\left( v \right),{p^*}\left( v \right),\forall v} \right\}$ into the primal problem \eqref{Problem2}, we have the following optimization problem:
\begin{subequations}\label{Problem3}
\begin{align}
&\mathop {\max }\limits_{\left\{ {\alpha \left( v \right)} \right\}} \;{{\mathbb{E}}_v}\left[ {\alpha \left( v \right)\left\{ {\rho^* \left( v \right){S_s^*}\left( v \right) + \left( {1 - \rho^* \left( v \right)} \right){S_b^*}\left( v \right)} \right\}} \right]\\
\label{C13}{\rm{s.t.}}\;\;&{{\mathbb{E}}_v}\left[ {\alpha \left( v \right){{\log }_2}\left( {1 + \frac{{{P_n}{{\left| {{h_n}\left( v \right)} \right|}^2}}}{{{p^*}\left( v \right){{\left| {{h_f}\left( v \right)} \right|}^2} + {\sigma ^2}}}} \right) + \left( {1 - \alpha \left( v \right)} \right){{\log }_2}\left( {1 + \frac{{{P_n}{{\left| {{h_n}\left( v \right)} \right|}^2}}}{{{\sigma ^2}}}} \right)} \right] \ge \overline R ,\\
\label{C23}&{{\mathbb{E}}_v}\left[ {{\alpha }\left( v \right)}{p^*\left( v \right)} \right] \le \overline P ,\\
\label{C33}&0 \le \alpha \left( v \right) \le 1, \forall v,
\end{align}
\end{subequations}
where ${S_s^*}\left( v \right) = \frac{I}{{KL}}{\widetilde \varepsilon _K}\left( {\frac{{p^*\left( v \right){{\left| {{h_f}\left( v \right)} \right|}^2}}}{{{\sigma ^2}}}} \right){\mathfrak{1}}\left( {{{\widetilde \varepsilon }_K} \ge \overline \varepsilon } \right)$ and ${S_b^*}\left( v \right) = \frac{I}{{\mu L}}{\log _2}\left( {1 + \frac{{p^*\left( v \right){{\left| {{h_f}\left( v \right)} \right|}^2}}}{{{\sigma ^2}}}} \right)$. It can be observed that problem \eqref{Problem3} is an LP problem with respect to ${\left\{ {\alpha \left( v \right)} \right\}}$, which can be solved by using standard convex optimization tools such as CVX~\cite{cvx}. The details of the procedures for optimally solving problem \eqref{Problem2} are summarized in \textbf{Algorithm 1}. 
\begin{algorithm}[!h]\label{method1}
\caption{Algorithm for Optimally Solving Problem \eqref{Problem2}} 
 \hspace*{0.02in}
\hspace*{0.02in} {Initialize an ellipsoid ${\mathcal{E}}\left( {\left( {\beta ,\delta } \right) ,{\mathbf{A}}} \right)$ containing $\left( {\beta^* ,\delta^* } \right) $, where $\left( {\beta ,\delta } \right)$ is the center point of  ${\mathcal{E}}\left( {\left( {\beta ,\delta } \right) ,{\mathbf{A}}} \right)$ and the positive definite matrix ${\mathbf{A}}$ characterizes the size of  ${\mathcal{E}}\left( {\left( {\beta ,\delta } \right) ,{\mathbf{A}}} \right)$.}\\
\vspace{-0.4cm}
\begin{algorithmic}[1]
\STATE {\bf Repeat}
\STATE \quad Obtain $\left\{ {{\rho ^*}\left( v \right),{\alpha ^*}\left( v \right),{p^*}\left( v \right)} \right\}$ by employing \eqref{roo}, and obtain ${g_2}\left( {\beta ,\delta } \right)$ under given $\left( {\beta ,\delta } \right)$.
\STATE \quad Compute the subgradients of ${g_2}\left( {\beta ,\delta } \right)$ using \eqref{sub-gradient}, and update $\left( {\beta ,\delta } \right)$ via the ellipsoid method.
\STATE {\bf Until} $\left( {\beta ,\delta } \right)$ converge with a prescribed accuracy.
\STATE Set $\left( {\beta^* ,\delta^* } \right)  \leftarrow  \left( {\beta ,\delta } \right)$.
\STATE Obtain $\left\{ {{\rho ^*}\left( v \right),{\alpha ^*}\left( v \right),{p^*}\left( v \right)} \right\}$ by employing \eqref{roo} and solving problem \eqref{Problem3}.
\end{algorithmic}
\end{algorithm}

\section{Numerical Results for the Proposed Opportunistic Scheme}
In this section, we provide numerical results to validate the effectiveness of the proposed opportunistic SemCom and BitCom scheme for NOMA. Unless stated otherwise, the adopted system parameters are the same with those in Section \ref{PC}(2). Specifically, the small-scale fading of the N-user-AP and F-user-AP channels at each fading state is assumed to be independent and identically distributed Rayleigh fading. The transmit power employed at the N-user is fixed at $P_n=1$ W. To verify the effectiveness of the proposed opportunistic SemCom and BitCom scheme, we consider the following two baseline schemes.
\begin{itemize}
  \item \textbf{SemCom-only Scheme}: In this case, we assume that the F-user uploads information to the AP via NOMA only relying on SemCom. The resultant semantic rate maximization problem over fading channel in Scenarios I and II can be solved by fixing $\left\{ {\rho \left( v \right) = 1,\forall v} \right\}$ in problems \eqref{Problem1} and \eqref{Problem2}.
  \item  \textbf{BitCom-only Scheme}: In this case, we assume that the F-user uploads information to the AP via NOMA only relying on BitCom. Similarly, the resultant equivalent semantic rate maximization problems in the two scenarios can be solved by fixing $\left\{ {\rho \left( v \right) = 0,\forall v} \right\}$ in problems \eqref{Problem1} and \eqref{Problem2}. 
\end{itemize}
\subsection{Scenario I: On-off Power Control and Time Scheduling}
In this subsection, we study the achieved performance in Scenario I, where the power control and time scheduling of the F-user in NOMA can be only adjusted in an on-off manner.
\subsubsection{Ergodic Semantic Rate versus $\overline R$} 
\begin{figure}[!t]
  \centering
  \begin{minipage}[t]{0.48\linewidth}
  \includegraphics[width=3.3in]{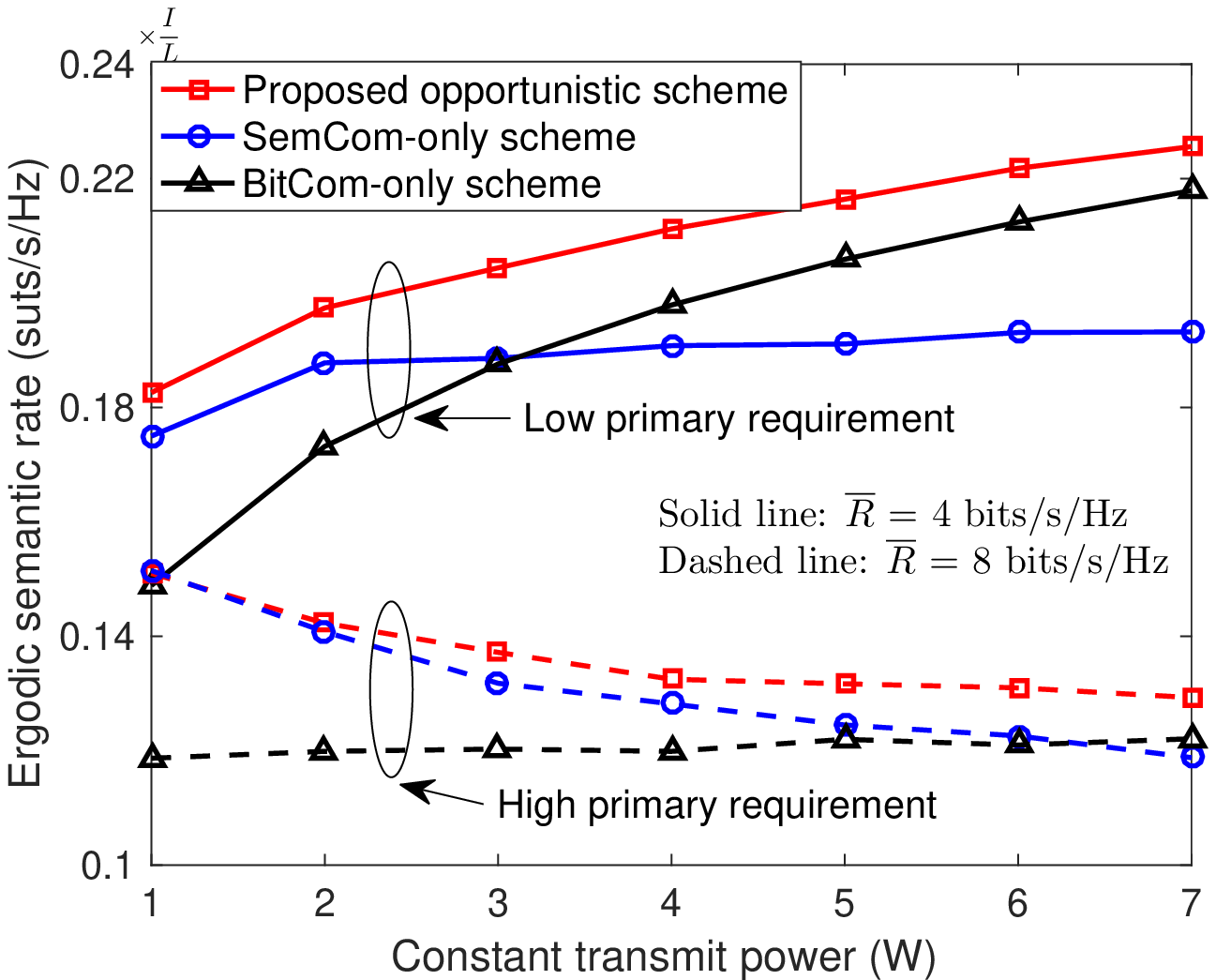}\\
  \caption{The achieved ergodic semantic rate of the F-user versus $\overline R$ in Scenario I.}\label{S_v_R1}
  \end{minipage}
  \quad
  \begin{minipage}[t]{0.48\linewidth}
      \includegraphics[width=3.3in]{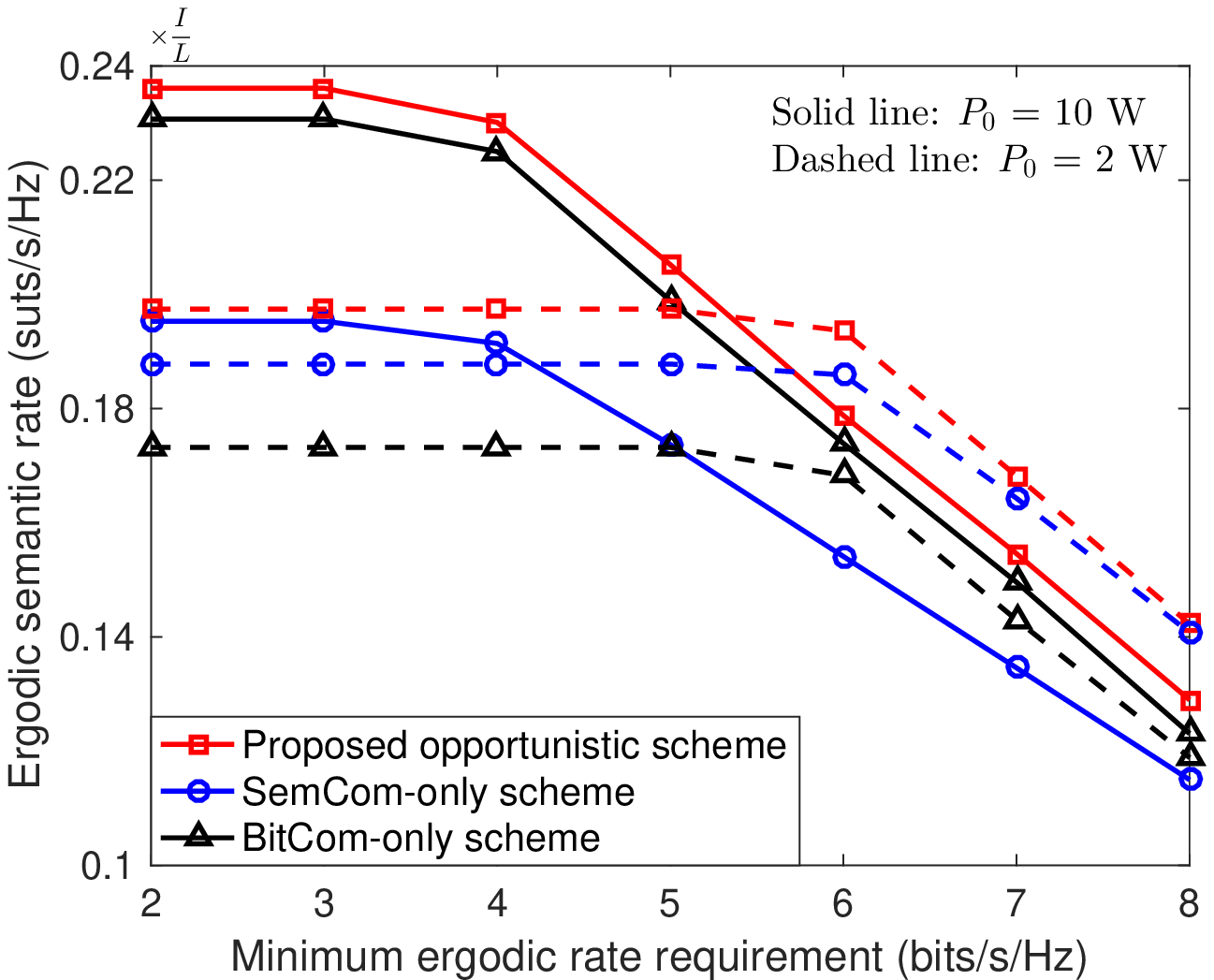}\\
  \caption{The achieved ergodic semantic rate of the F-user versus $P_0$ in Scenario I.}\label{S_v_P1}
  \end{minipage}
\end{figure}

In Fig. \ref{S_v_R1}, we investigate the ergodic semantic rate achieved at the F-user, ${{{\mathbb{E}}_v}\left[ {S\left( v \right)} \right]}$, versus the minimum ergodic bit rate required at the N-user, $\overline R$. For the constant transmit power that can be employed at the F-user, we consider two cases: (1) $P_0=2$ W and (2) $P_n=10$ W. As can be observed from Fig. \ref{S_v_R1}, the ergodic semantic rate obtained by all schemes decreases as $\overline R$ increases. This is because a higher $\overline R$ leads to a less interference level can be tolerant when the AP decodes the primary N-user's signal, see \eqref{B_R_N}, thus resulting in a lower ergodic semantic rate of the F-user. It can be also observed that the proposed opportunistic SemCom and BitCom scheme always achieves the best performance in the two cases. This is indeed expected since the proposed opportunistic scheme provides more degrees-of-freedom (DoFs) in communication design. Instead of merely relying on one specific communication method, the proposed opportunistic scheme enables the F-user to select the suitable communication method in each fading state, thus having a higher probability to be admitted in the primary N-user's resource block. Moreover, as seen from Fig. \ref{S_v_R1}, the SemCom-only scheme outperforms the BitCom-only scheme for the case of $P_0=2$ W, but performs significantly worse than the BitCom-only scheme for the case of $P_0=10$ W. This is also consistent with our theoretical analysis and numerical examples in Section \ref{PC}, i.e., SemCom and BitCom are generally preferred to be employed in the high SNR and low SNR regimes, respectively. It also shows that there is only a small gap between the proposed opportunistic scheme and the BitCom-only scheme for $P_0=10$ W, which implies that the F-user mainly employs BitCom in this case. The communication mode preferred by the F-user in different cases will be further verified in the following results, see Fig. \ref{time}. The above results confirm the effectiveness of the proposed opportunistic SemCom and BitCom scheme.
\subsubsection{Ergodic Semantic Rate versus $P_0$} In Fig. \ref{S_v_P1}, we investigate the ergodic semantic rate achieved at the F-user, ${{{\mathbb{E}}_v}\left[ {S\left( v \right)} \right]}$, versus the F-user's constant transmit power, $P_0$. For the minimum ergodic bit rate required at the N-user, we consider two cases: (1) $\overline R = 4$ bits/s/Hz and (2) $\overline R = 8$ bits/s/Hz. As illustrated in Fig. \ref{S_v_P1}, the proposed opportunistic scheme achieves the best performance in both two cases. It can be observed that when $P_0$ increases, the ergodic semantic rate obtained by all schemes increases for the case of $\overline R = 4$ bits/s/Hz, while the ergodic semantic rate obtained by the proposed opportunistic scheme and SemCom-only scheme decreases for the case of $\overline R = 8$ bits/s/Hz. The reasons behind this can be explained as follows. When the N-user has a small ergodic bit rate requirement (e.g., $\overline R = 4$ bits/s/Hz), the interference level that can be tolerant when the AP decodes the N-user's bit signal is high. This allows the F-user to be admitted via NOMA and the corresponding permitted transmit power of the F-user is also high. Therefore, increasing $P_0$ in this case benefits the ergodic semantic rate achieved by the F-user. Moreover, it can be observed that the performance gain achieved by increasing $P_0$ in the proposed opportunistic scheme and BitCom-only scheme is more pronounced than that in the SemCom-only scheme. This is because the semantic rate is upper-bounded in the high SNR regime, see \eqref{s_NOMA}. When the achieved semantic rate approaches its upper bound, further increasing the transmit power will not be significantly beneficial as that in BitCom. This also underscores the importance of the proposed opportunistic SemCom and BitCom scheme for exploiting the benefits of the two technologies. However, when the N-user has a relatively high ergodic bit rate requirement (e.g., $\overline R = 8$ bits/s/Hz), the interference level that can be tolerated when decoding the N-user's bit signal is low. In this case, increasing $P_0$ will not benefit the performance of the F-user since it reduce the number of fading states where the F-user can be admitted via NOMA given the resulting high interference. Therefore, the ergodic semantic rate of the F-user decreases due to the limited transmission time over fading channels.    
\subsubsection{Communication Methods Employed at the F-user} 
\begin{figure}[!h]
  \centering
  \includegraphics[width=3.5in]{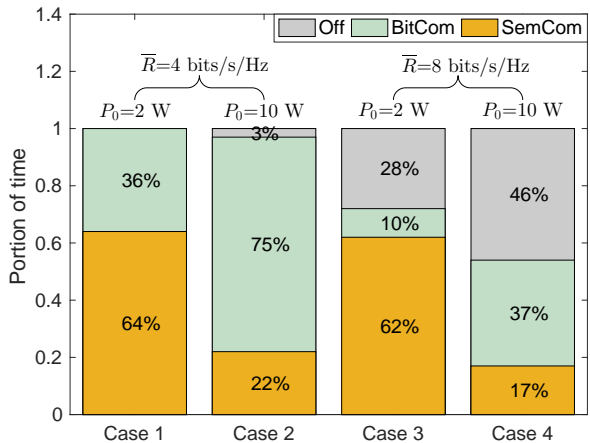}\\
  \caption{The time portion of three working modes at the F-user under different cases in Scenario I.}\label{time}
\end{figure}

In Fig. \ref{time}, we plot the time portion of ``Off'', ``BitCom'', and ``SemCom'' modes adopted at the F-user when using the proposed opportunistic scheme. We consider four cases: (1) $\overline R = 4$ bits/s/Hz, $P_0=2$ W; (2) $\overline R = 4$ bits/s/Hz, $P_0=10$ W; (3) $\overline R = 8$ bits/s/Hz, $P_0=2$ W; and (4) $\overline R = 8$ bits/s/Hz, $P_0=10$ W. It can be observed that a small $\overline R$ allows the F-user to be admitted in most fading states, while for a large $\overline R$, the F-user has to turn off in a considerable number of fading states to reduce the interference caused to the N-user, thus yielding a limited ergodic semantic rate as shown in Fig. \ref{S_v_R1}. Regarding the employed communication method, it can be observed that SemCom servers as the main method when $P_0=2$ W and BitCom becomes dominated when $P_0=10$ W. For $\overline R = 8$ bits/s/Hz, the active time of the F-user with $P_0=10$ W is significantly reduced compared to that with $P_0=2$ W. This is because, the N-user in this case is sensitive to the interference, which prevents the F-user from being admitted, especially when $P_0$ is large. Therefore, a higher $P_0$ will make the F-user achieve a lower ergodic semantic rate, as shown in Fig. \ref{S_v_P1}.
\subsection{Scenario II: Continuous Power Control and Time Scheduling}
In this subsection, we continue to study the achieved performance in Scenario II, where the power control and time scheduling of the F-user in NOMA can be continuously adjusted.
\subsubsection{Ergodic Semantic Rate versus $\overline R$} 
\begin{figure}[!t]
  \centering
  \begin{minipage}[t]{0.48\linewidth}
  \includegraphics[width=3.3in]{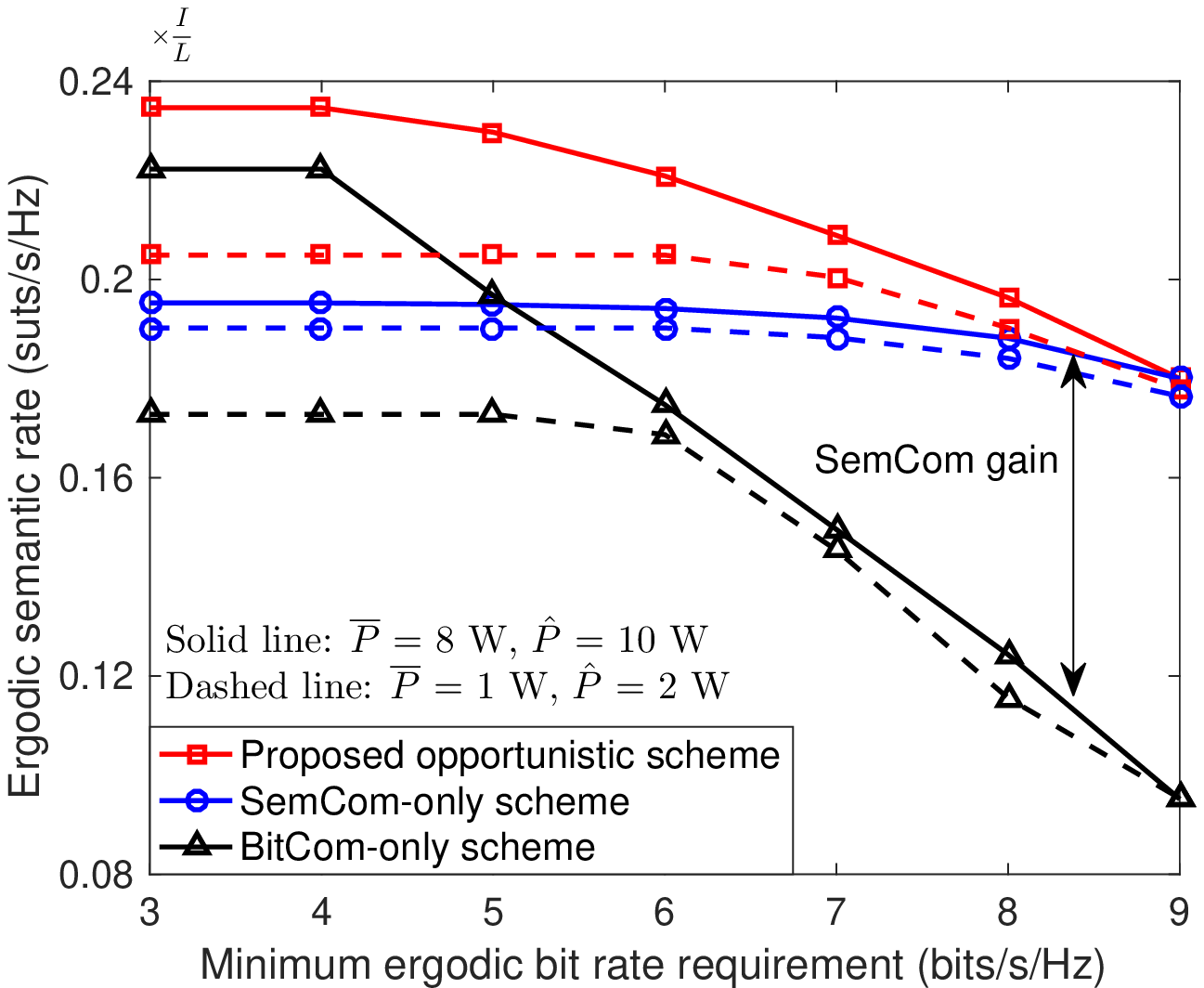}\\
  \caption{The achieved ergodic semantic rate of the F-user versus $\overline R$ in Scenario II.}\label{S_v_R2}
  \end{minipage}
  \quad
  \begin{minipage}[t]{0.48\linewidth}
      \includegraphics[width=3.3in]{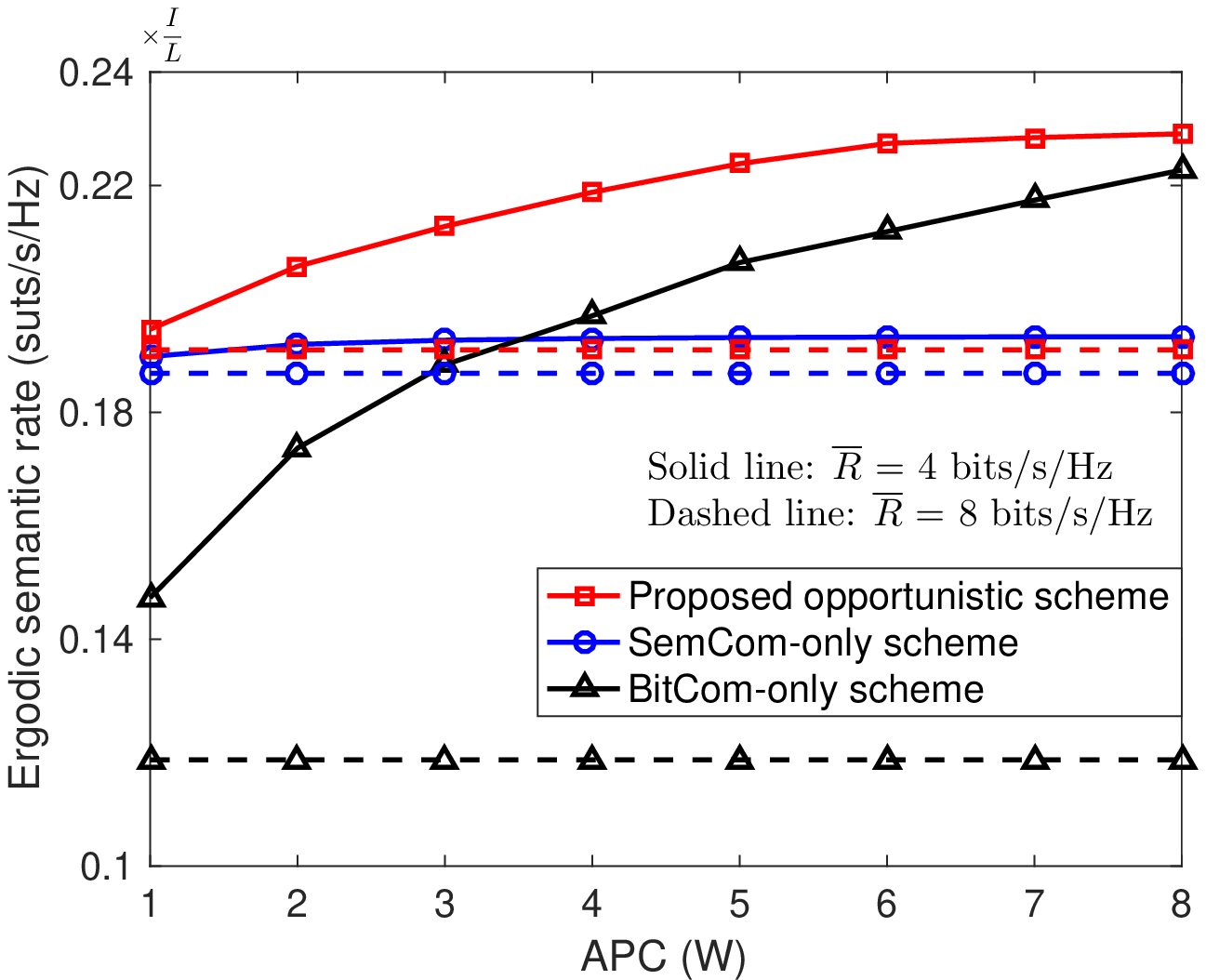}\\
  \caption{The achieved ergodic semantic rate of the F-user versus the APC, $\overline P$, in Scenario II.}\label{S_v_P2}
  \end{minipage}
\end{figure}

In Fig. \ref{S_v_R2}, we investigate the ergodic semantic rate achieved at the F-user, ${{{\mathbb{E}}_v}\left[ {S\left( v \right)} \right]}$, versus the minimum ergodic bit rate required at the N-user, $\overline R$, in Scenario II. For the PPC and APC, we consider two cases: (1) $\overline P = 1$ W, $\hat P = 2$ W; and (2) $\overline P = 8$ W, $\hat P = 10$ W. As depicted in Fig. \ref{S_v_R2}, the ergodic semantic rate obtained by all the schemes with a high transmit power budget is larger than that with a small transmit power budget. This is because the continuous power control make the setup of $\overline P = 8$ W and $\hat P = 10$ W include the setup of $\overline P = 1$ W, $\hat P = 2$ W as a special case. It can be also observed that in the small $\overline R$ regime (i.e., the F-user can be admitted using a high transmit power), the performance gain via increasing the transmit power budget in the proposed opportunistic scheme and the BitCom-only scheme is much more significant than that in the SemCom-only scheme. However, in the high $\overline R$ regime, the performance of the BitCom-only scheme is significantly degraded as compared with the other two schemes. This is because the N-user in this case is sensitive to the interference, only a small transmit power is permitted to the F-user if admitted. In such a case, where the feasible transmit power of the F-user is strictly capped, the employment of SemCom ensures that the F-user in NOMA can still achieve a considerable high communication performance as compared to the BitCom-only scheme. As a result, it can be seen that the performance gap between the proposed opportunistic scheme and the SemCom-only scheme is negligible in the high $\overline R$ regime since SemCom is the main communication method in this case. This result shows again that employing SemCom can have a salient performance gain for NOMA, i.e., alleviating the ``early-late' rate disparity issue. 
\subsubsection{Ergodic Semantic Rate versus $\overline P$} In Fig. \ref{S_v_P2}, we investigate the ergodic semantic rate achieved at the F-user, ${{{\mathbb{E}}_v}\left[ {S\left( v \right)} \right]}$, versus the APC, $\overline P$, in Scenario II, where the PPC is $\hat P = 10$W. For the minimum ergodic bit rate required at the N-user, we consider two cases: (1) $\overline R = 4$ bits/s/Hz and (2) $\overline R = 8$ bits/s/Hz. It can be observed that the ergodic semantic rate obtained by all schemes increases for the case of $\overline R = 4$ bits/s/Hz, but remains almost unchanged for the case of $\overline R = 8$ bits/s/Hz. This is because in the low $\overline R$ regime, increasing $\overline P$ allows the F-user to be admitted in NOMA using a higher transmit power to improve its communication performance. Specifically, the SemCom-only scheme can achieve better performance than the BitCom-only scheme for small $\overline P$ and becomes the worst scheme for high $\overline P$. This is also consistent with our theoretical analysis and numerical examples in Section \ref{PC}. The proposed opportunistic scheme always achieves the best performance. However, for $\overline R = 8$ bits/s/Hz, the permitted transmit power of the F-user is strictly capped when being admitted in NOMA. In this case, the APC becomes dummy and increasing $\overline P$ will not improve the performance of the F-user in all schemes. 

\subsubsection{The Impact of Power Control and Time Scheduling} Finally, we investigate the impact of the adjustment feature of power control, ${\left\{ {p \left( v \right)} \right\}}$, and time scheduling, ${\left\{ {\alpha \left( v \right)} \right\}}$, on the achieved communication performance over fading channels. Besides the continuous resource management in power control and time scheduling, we consider the following three baseline schemes.
\begin{itemize}
  \item \textbf{Continuous power control and on-off time scheduling}: In this scheme, we assume that the employed transmit power can be continuously adjusted subject to the PPC and APC, while the time scheduling can only be adjusted in an on-off manner. 
  \item \textbf{On-off power control and continuous time scheduling}: In this scheme, we assume that the time scheduling can be continuously adjusted, while the employed transmit power can only be zero and $\hat P$ subject to the APC. 
  \item \textbf{On-off power control and time scheduling}: In this scheme, we assume that both the transmit power and time scheduling can only be adjusted in an on-off manner, as assumed in Scenario I, but subject to the APC.
\end{itemize}
\begin{figure}[!b]
  \centering
  \includegraphics[width=3.5in]{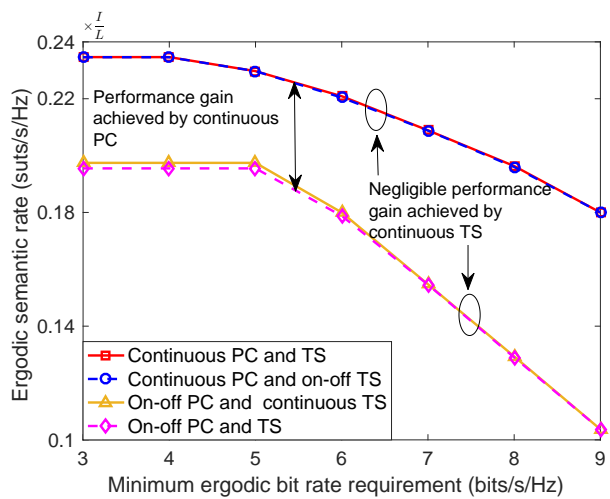}\\
  \caption{The performance comparison among different resource management schemes, where ``PC'' and ``TS'' refer to power control and time scheduling, respectively.}\label{Compare}
\end{figure}
In Fig. \ref{Compare}, we present the ergodic semantic rate obtained by different resource management schemes versus for the minimum ergodic bit rate required at the N-user, $\overline R$. Here, we set $\overline P = 8$ W for APC and $\hat P = 10$ W for PPC. It is interesting to observe that, regardless of what kind of adjusting strategies assumed for the time scheduling, there is a noticeable performance gap between the schemes using continuous power control and those using on-off power control. This is because the continuous power control enables the F-user to well manage the caused interference, thus having a high probability to be admitted in NOMA and achieving a higher ergodic semantic rate. It can be also observed that there is only a small performance gap between the continuous and on-off time scheduling if the on-off power control is used, which further becomes negligible if the continuous power control is employed. The above results underscore the importance of continuous power control for enabling the F-user to carry out interference management, thus being admitted in NOMA. It also indicates that in practical long-term transmission, employing the simple on-off time scheduling at the F-user is sufficient. In other words, the continuous power control at the F-user is more efficient than the continuous time scheduling for managing the interference over fading channels. 

\section{Conclusions} 
The exploitation of SemCom in NOMA has been investigated in this paper. In particular, a novel semantics-empowered two-user uplink NOMA framework was conceived, where a primary N-user and a secondary F-user communicate with the AP using BitCom and SemCom, respectively. To management of the interference caused to the primary N-user, the total transmission is partially allocated for admitting the secondary F-user in NOMA. By comparing the achieved SvB rate region of the proposed semantics-empowered NOMA framework with the equivalent one of the conventional BitCom-based NOMA, it reveals that SemCom has the great potentials of further improving the secondary F-user's performance without degrading the primary N-user's performance, thus alleviating the ``early-late'' rate disparity issue for NOMA. However, it also shows that SemCom may perform worse than BitCom in specific conditions. Motivated by this result, the proposed semantics-empowered NOMA framework was further investigated over fading channels. An opportunistic SemCom and BitCom scheme was proposed for enabling the secondary F-user to exploit the benefits of two technologies when being admitted in NOMA at each fading state. Considering both the on-off and continuous resource management scenarios, the optimal communication policy was derived for maximizing the ergodic semantic rate of the secondary F-user while satisfying the minimum ergodic bit rate requirement of the primary N-user. The presented numerical results confirmed the effectiveness of the proposed opportunistic scheme over other baseline schemes. It revealed that compared to BitCom, employing SemCom in NOMA can effectively guarantee the performance of the secondary F-user especially when the communication requirement of the primary N-user is high. The obtained results also revealed that, compared to time scheduling, the continuous power control is more essential for ensuring the communication performance achieved by the secondary F-user over fading channels.

\bibliographystyle{IEEEtran}
\begin{spacing}{1.0}
\bibliography{mybib}
\end{spacing}
\end{document}